%% file: main.tex
\documentclass[%
reprint,
amsmath,
amssymb,
aps,
prl
]{revtex4-2}

\usepackage{graphicx}
\usepackage{bm}
\usepackage{physics}
\usepackage{amsthm}
\usepackage{color}
\usepackage{titlesec}
\usepackage[normalem]{ulem}
\usepackage{enumitem}
\usepackage{pifont}
\usepackage{cancel}
\usepackage[caption=false]{subfig}
\usepackage[colorlinks=true, citecolor=blue, urlcolor=blue, linkcolor=blue, pdfencoding=auto, psdextra]{hyperref}
\usepackage[all]{hypcap}
\usepackage{mathtools}
\usepackage{tikz}

\newcommand\dual[1]{\overline{#1}}

\newcommand\dualmathcal[1]{\mathcal{\mskip.9\thinmuskip\overline{\mskip-.9\thinmuskip {#1} \mskip-.1\thinmuskip}\mskip.1\thinmuskip}}

\newcommand{\rainbow}{%
  \begin{tikzpicture}[scale=0.5]
    \draw (0,0) arc (0:180:5mm);
    \draw (-1.5mm,0) arc (0:180:3.5mm);
    \draw (-3mm,0) arc (0:180:2mm);
    \filldraw (0,0) circle (1pt);
    \filldraw (-1.5mm,0) circle (1pt);
    \filldraw (-3mm,0) circle (1pt);

    \filldraw (-7mm,0) circle (1pt);
    \filldraw (-8.5mm,0) circle (1pt);
    \filldraw (-10mm,0) circle (1pt);
  \end{tikzpicture}%
}

\newif\ifnosupp
\newif\ifnocrosslinks
\IfFileExists{./supp.tex}{}{\nosupptrue\nocrosslinkstrue}

\ifnocrosslinks
  \newcommand\suppref[1]{\ref*{#1}}
  \newcommand\suppcite[1]{{\protect\NoHyper\cite{#1}\protect\endNoHyper}}
\else
  \newcommand\suppref[1]{\ref{#1}}
  \newcommand\suppcite[1]{\cite{#1}}
\fi

\titleformat{\section}[runin]
            {\normalfont\itshape}{\thesection}{1em}{}[---]
\titlespacing*{\section}
              {1em}
              {0.3em}
              {0.0em}

\titleformat{\subfigure}[leftmargin]{}{}{0pt}{}
              
\setlist{nosep}
\newtheorem{theorem}{Theorem}
\newtheorem{lemma}[theorem]{Lemma}

\makeatletter
\newcommand{\dummylabel}[2]{\def\@currentlabel{#2}\label{#1}}
\def\maketitle{
\@author@finish
\title@column\titleblock@produce
\suppressfloats[t]}
\makeatother

\begin{document}
\title{Volume-entangled exact scar states in the PXP and related models in any dimension}
\author{Andrew~N.~Ivanov and Olexei~I.~Motrunich}
\affiliation{Department of Physics and Institute for Quantum Information and Matter, California Institute of Technology, Pasadena, California 91125, USA}

\date{\today}

\begin{abstract}
  In this Letter, we report first exact volume-entangled Einstein-Podolsky-Rosen--type scar states hosted by PXP and related Hamiltonians corresponding to various geometric configurations of Rydberg-blockaded atom systems, including the most extensively studied ones such as the chain with periodic boundary conditions (PBCs) and square lattice.
We start by introducing a new zero-energy eigenstate of the PBC chain and proceed by generalizing it to a wide variety of geometries and Hamiltonians.
We point out the potential experimental relevance of our states by providing a protocol for their preparation on near-term Rydberg quantum devices, which relies only on strictly local measurements and evolution under native Hamiltonians.
We also demonstrate the utility of these states for the study of quantum dynamics by describing a protocol for measuring infinite-temperature out-of-time-order correlator functions.
\end{abstract}

\maketitle
\phantomsection
\section*{Introduction}
\addcontentsline{toc}{section}{Introduction}
The observation of unusually slow thermalization and unexpected many-body revivals during a quench from a N\'{e}el-type state in the pioneering Rydberg atom experiment~\cite{Bernien_2017} ignited interest in the so-called PXP model~\cite{Lesanovsky_2012}, which is an idealized description of Rydberg atomic systems in the nearest-neighbor blockade regime.
The attribution of this atypical dynamics to the special ``scar'' eigenstates weakly violating the eigenstate thermalization hypothesis (ETH) in the spectrum of the one-dimensional (1D) PXP Hamiltonian~\cite{Turner_2018weak, Turner_2018quantum} opened the field of quantum many-body scars (QMBSs)~\cite{Shiraishi_2017, Moudgalya_2018exact, Moudgalya_2018entanglement}.
Various perspectives on the mechanisms underlying these states have been put forth~\cite{Turner_2018weak, Turner_2018quantum, Khemani_2019, Lin_2019, Choi_2019, Surace_2020, Iadecola_2019, Shiraishi_2019, Omiya_2023quantum, Omiya_2023fractionalization, Ljubotina_2023, giudici2023unraveling}, yet many open questions prompting active research of PXP-type models still remain (for general QMBS reviews see~\cite{Serbyn_2021, Papic_2022, Moudgalya_2022, Chandran_2023, Moudgalya_2023exhaustive}).

\begin{figure}
  \includegraphics[width=\columnwidth]{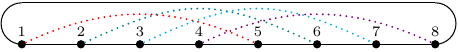}
  \caption{State $\ket{\Lambda}$ on an $N=2L=8$ PBC chain.
  Dotted lines indicate the pairing pattern of perfectly correlated spins.}
  \label{fig:pbcchain}
\end{figure}

This Letter enriches our understanding of PXP-type Hamiltonians --- which despite their apparent simplicity are nonintegrable and chaotic based on the level statistics --- by discovering exact Einstein-Podolsky-Rosen (EPR) type volume-entangled eigenstates in a variety of geometries, starting from the 1D chain with periodic boundary conditions (PBCs) shown in Fig.~\ref{fig:pbcchain}.
We expect our findings to be relevant to the broad field of quantum information processing, particularly for tasks involving production and utilization of large-scale entanglement.
Unlike the previous exact results (the eigenstates of the 1D chain found in Ref.~\cite{Lin_2019} and states with valence bond solid orders in two-dimensional models~\cite{Lin_2020, Michailidis_2020}), which for standard bipartitions all had area law scaling of entanglement, our new states exhibit volume-law entanglement growth and thus highlight a previously unknown mechanism of weak ETH violation in PXP-type Hamiltonians.
Appearing locally as infinite-temperature Gibbs ensembles, these states provide a rare analytical glimpse into the emergence of statistical mechanics in realistic closed nonintegrable quantum systems.
Notably, the translational-symmetry-breaking PBC state found in Ref.~\cite{Lin_2019}, its four variants for open boundary conditions (OBCs), and our newly introduced state are the only known exact eigenstates of the 1D PXP chains.

This Letter is organized as follows.
First, we introduce and prove a new $E=0$ eigenstate of the 1D PXP chain with PBCs. 
This state is then compared to other constructions, and generalized to many experimentally relevant geometries.
Finally, to showcase the applicability of our results to the study of quantum many-body dynamics (or demonstrations of precise quantum control) on analog Rydberg quantum devices~\cite{Labuhn_2016, Browaeys_2020, Altman_2021, Ebadi_2022, Bluvstein_2023, Madjarov_2020, Choi_2023, Mark_2023, shaw2023benchmarking, anand2024dualspecies}, we sketch several experimental proposals, keeping in mind that further studies are needed to assess their practical feasibility.

\section*{Eigenstate of the 1D PXP chain with PBCs}
\addcontentsline{toc}{section}{Eigenstate of the PXP chain}
Consider a state on a spin-1/2 chain of size $N = 2L$ with PBCs defined as
\begin{equation}
  \label{eq:lambda}
  \ket{\Lambda} =
  \frac{1}{\sqrt{\chi}}\sum_{f \in
    \mathcal{F}^{\text{PBC}}_L}(-1)^{|f|}\ket{f}_{1,\ldots,L}\otimes\ket{f}_{L+1,\ldots,2L},
\end{equation}
where $\mathcal{F}^{\text{PBC}}_L$ is the set of bitstrings defining the nearest neighbor Rydberg-blockaded subspace for chains of $L$ spins with PBCs, $|f|$ denotes the Hamming weight of the bitstring $f$ (i.e., the number of 1's), and, in terms of the Fibonacci numbers $F_n$, $\chi = \left|\mathcal{F}^{\text{PBC}}_L\right| = F_{L-1} + F_{L+1}$ is the dimension of the Rydberg-blockaded Hilbert space of a PBC chain of size $L$ \cite{Turner_2018weak, Turner_2018quantum}.
Figure ~\ref{fig:pbcchain} illustrates the correlation (spin pairing) pattern in $\ket{\Lambda}$.

By construction, each $\ket{f}\otimes\ket{f}$ in Eq.~(\ref{eq:lambda}) satisfies the Rydberg blockade on the full PBC chain. The state $\ket{\Lambda}$ is invariant (eigenvalue 1) under the spectral reflection symmetry operator $\mathcal{C} = \prod_{i=1}^NZ_i$ (where $Z_i \equiv \ketbra{0}_i - \ketbra{1}_i$)~\footnote{Operator $\mathcal{C}$ is sometimes called the ``particle-hole'' symmetry (and denoted $\mathcal{C}_\text{ph}$) because it anticommutes with the Hamiltonian and thus relates positive and negative energy eigenstates.}, lattice translation by one site, inversion, and pseudo-local $\text{SWAP}_{i, i+L}$ and $Z_{i} Z_{i+L}$ gates. Furthermore,
\begin{theorem}
  \label{thrm:lambda}
  $\ket{\Lambda}$ is an exact zero energy eigenstate of the Hamiltonian
  \begin{equation}
    \label{eq:hpxpring}
    H_\mathrm{PXP}(N) = \sum_{i=1}^N P_{i-1} X_i P_{i+1}, \quad (i + N \equiv i),
  \end{equation}
  where $P_i \equiv \ketbra{0}{0}_i$ and $X_i \equiv \ketbra{1}{0}_i + \ketbra{0}{1}_i$.
\begin{proof}
For any term in Eq.~(\ref{eq:lambda}), the neighborhoods of spins $i \in \{1,\dots,L\}$ and $i+L$ are identical in the full system $[1,\dots,2L]$, and half-systems $[1,\dots,L]$ and $[L+1,\dots,2L]$ with assumed PBCs.
This means that the action of $H_\text{PXP}(N)$ on $\ket{\Lambda}$ must be equivalent to the action of the ``disconnected-halves'' (DH) Hamiltonian --- i.e., $H_\text{PXP}(N)\ket{\Lambda} = H_{\text{DH}}(L)\ket{\Lambda} = 0$ --- written as a sum of two decoupled PXP Hamiltonians with half-system PBCs:
\begin{align}
  \label{eq:heff}
  & H_\text{DH}(L) = H_\text{PXP}^\text{left}(L)_{1,\ldots,L}
  + H_\text{PXP}^\text{right}(L)_{L+1,\ldots,2L}.
\end{align}
It then remains to show that $H_\text{PXP}^\text{left}(L)\ket{\Lambda} =-H_\text{PXP}^\text{right}(L)\ket{\Lambda}$. Individually,
\begin{subequations}  
  \begin{align}
    \label{eq:hllambda}
    &H_\text{PXP}^\text{left}(L)\!\ket{\Lambda} = \chi^{\scriptscriptstyle-\frac{1}{2}}\!\sum_{\mathclap{f \in
      \mathcal{F}^{\text{PBC}}_L}}(-1)^{|f|}[H_\text{PXP}(L)\ket{f}]\!\otimes\!\ket{f}, \\
    \label{eq:hrlambda}
    &H_\text{PXP}^\text{right}(L)\!\ket{\Lambda} = \chi^{\scriptscriptstyle-\frac{1}{2}}\!\sum_{\mathclap{f \in \mathcal{F}^{\text{PBC}}_L}}(-1)^{|f|}\ket{f}\!\otimes\![H_\text{PXP}(L)\ket{f}].
  \end{align}
\end{subequations}
Multiplying Eqs.~(\ref{eq:hllambda}) and (\ref{eq:hrlambda}) by resolutions of identity
$\left(\sum_{g \in \mathcal{F}^{\text{PBC}}_L}\ketbra{g}{g}
\right)_{1,\ldots,L}$ and
$\left(\sum_{g \in
    \mathcal{F}^{\text{PBC}}_L}\ketbra{g}{g}\right)_{L+1,\ldots,2L}$,
respectively, gives \footnote{It is acceptable for the terms in the resolutions of identity to satisfy the half-system Rydberg blockade for PBCs and not OBCs because, by construction, $H_\text{DH}(L)$ acting on $\ket{\Lambda}$ does not produce any components that violate the PBC Rydberg blockade for any half of the system.}
\begin{subequations}
  \begin{align}
    H_\text{PXP}^\text{left}(L)\ket{\Lambda} &= \chi^{\scriptscriptstyle-\frac{1}{2}}\!
    \sum_{\mathclap{f,g \in \mathcal{F}^{\text{PBC}}_L}}\gamma(f,g)\ket{g}\otimes\ket{f},\\
    H_\text{PXP}^\text{right}(L)\ket{\Lambda} &= \chi^{\scriptscriptstyle-\frac{1}{2}}\!
    \sum_{\mathclap{f,g \in \mathcal{F}^{\text{PBC}}_L}}\gamma(f,g)\ket{f}\otimes\ket{g},
  \end{align}
\end{subequations}
where $\gamma(f,g) = (-1)^{|f|}\mel{g}{H_\text{PXP}(L)}{f} \in \mathbb{R}$. Since $\mel{g}{H_\text{PXP}(L)}{f} \neq 0$ only if $|f| \neq |g| \mod 2$, it follows that $\gamma(f, g) = -\gamma(g, f)$ and, therefore, $H_\text{PXP}^\text{left}(L)\ket{\Lambda} =-H_\text{PXP}^\text{right}(L)\ket{\Lambda}$.
\end{proof}
\end{theorem}

The same argument, with an appropriate substitution for $\mathcal{F}_L^{\text{PBC}}$ in Eq.~(\ref{eq:lambda}), holds for Hamiltonians whose terms have the form $A_{[i-m, \dots, i-1]} X_i B_{[i+1, \dots, i+n]}$, where $A$ and $B$ are diagonal in the computational basis operators with supports as indicated, in particular for all deformations and extensions (including longer-range blockades) of the PXP Hamiltonian in Refs.~\cite{Khemani_2019, Choi_2019, Karle2021arealaw, Surace_2021}.
Also, terms at different locations can come with different couplings as long as they are identical for sites $i$ and $i+L$ (see Sec.~\suppref{supp:proofalt} of \cite{supp}\nocite{Cirac_2021}\nocite{mcculloch2008infinitesizedensitymatrix} for alternative proof of Theorem~\ref{thrm:lambda} where this is more explicit).
For instance, if $L$ is even, Hamiltonians $H_o$ and $H_e$ defined by restricting the sum in Eq.~(\ref{eq:hpxpring}) to, respectively, odd and even sites both annihilate $\ket{\Lambda}$, which means this state is invariant under Floquet unitaries $U^{o/e}_\tau = e^{-iH_{o/e}\tau}$ used to define the Floquet PXP model in Refs.~\cite{Iadecola_2020, Rozon_2022, giudici2023unraveling}.
Generalizations discussed here also apply to states introduced in later sections.

Tracing over sites $L+1,\dots,2L$ in Eq.~(\ref{eq:lambda}) yields the maximally mixed Gibbs ensemble on a Rydberg-blockaded PBC chain of size $L$; therefore, all observables supported on a half-system have thermal expectation values.
The entanglement entropy of such \emph{standard} bipartition into two half-systems is $\log\chi \propto L$ (i.e., volume law).
In contrast, fine-tuned \emph{entanglement-minimizing} bipartitions that place both spins in each perfectly correlated spin pair into the same subsystem, while cutting as few physical bonds as possible (which for $\ket{\Lambda}$ amounts to four), achieve characteristic for QMBS (and highly unusual for states in the middle of the energy spectrum) area-law entropy scaling.
Also, consistent with other scar states (e.g., those in Ref.~\cite{Lin_2019}), $\ket{\Lambda}$ admits an exact finite-bond-dimension matrix product state (MPS) representation.
The entanglement structure of $\ket{\Lambda}$ and its MPS representation are discussed in Secs.~\suppref{supp:entstr} and \suppref{supp:mps} of~\cite{supp}, respectively.

\section*{Relation to other volume-entangled states}
\addcontentsline{toc}{section}{Relation to other volume-entangled states}
Among existing constructions on a doubled Hilbert space with cut-dependent volume- and area-law entropy scalings~\cite{Cottrell2019, Wildeboer_2022}, the rainbow scars~\cite{Langlett_2022} characterized by a rainbow-like pattern of nested Bell pairs with mirror symmetry (such as \rainbow) bear the closest resemblance to $\ket{\Lambda}$.
To gain a better insight into this connection, consider a DH Hamiltonian [cf.~Eq.~(\ref{eq:heff})]
\begin{equation}
  \label{eq:hambipart}
  H_\text{DH} = H_\mathcal{A} + H_\dualmathcal{A},
\end{equation}
whose real-valued (in the bitstring basis) terms act identically on identical systems $\mathcal{A}$ and $\dualmathcal{A}$, having the previously defined $\mathcal{C}$ as their spectral reflection symmetry \cite{Schecter_2018} operator (i.e., $\{H_{\mathcal{A}/\dualmathcal{A}}, \mathcal{C}_{\mathcal{A}/\dualmathcal{A}}\} = 0$) \footnote{Real-valuedness of the Hamiltonians in the bitstring basis (and, hence, of their eigenstates), and their spectral reflection symmetry under $\mathcal{C}$ are characteristic features for the PXP-type models being considered.}.
Note that we only stipulate that $\mathcal{A}$ and $\dualmathcal{A}$ are isomorphic, whereas the rainbow construction relies on a mirror symmetry between them and thus forbids translationally invariant (TI) states like our $\ket{\Lambda}$ (as well as its higher-dimensional generalizations to a variety of noteworthy systems to be introduced later).

The state
\begin{equation}
  \label{eq:lambdaeigen}
  \ket*{\widetilde\Lambda} = \frac{1}{\sqrt{\dim
      \mathcal{K}}}\sum_{n=1}^{\dim\mathcal{K}} \ket{\psi_n}_\mathcal{A}\otimes
  \mathcal{C}\ket{\psi_n}_\dualmathcal{A},
\end{equation}
where $\ket{\psi_n}$ are the real-valued eigenstates of $H_\mathcal{A}$ spanning a Krylov subspace $\mathcal{K}$, is a zero energy eigenstate of $H_\text{DH}$ since $H_\text{DH}\left(\ket{\psi_n}\otimes\mathcal{C}\ket{\psi_n}\right) = 0$ for $\forall n$.
Even though $\ket{\psi_n}$ are not known, we can write $\ket*{\widetilde\Lambda}$ in terms of a bitstring set $\mathcal{F}: \mathcal{K} = \operatorname{span}\{\ket{f}:f\in\mathcal{F}\}$.  Multiplying Eq.~(\ref{eq:lambdaeigen}) by $\mathbf{1} = \sum_{f,g}\ketbra{f}{f}_\mathcal{A}\otimes\ketbra{g}{g}_\dualmathcal{A}$, where $f, g \in \mathcal{F}$, we obtain [using $\mathcal{C}\ket{g} = (-1)^{|g|} \ket{g}$ and $\braket{g}{\psi_n} \in \mathbb{R}$]
\begin{equation}
  \label{eq:lambdagen}
  \ket*{\widetilde\Lambda}_\mathcal{F}=\frac{1}{\sqrt{\left|\mathcal{F}\right|}}\sum_{f \in \mathcal{F}}(-1)^{|f|}\ket{f}_\mathcal{A}\otimes\ket{f}_\dualmathcal{A},
\end{equation}
which matches Eq.~(\ref{eq:lambda}) for $\mathcal{F}= \mathcal{F}_L^{\text{PBC}}$ \footnote{It can be said that $\ket*{\widetilde\Lambda}$ is a purification of the maximally mixed ensemble on $\mathcal{K}$}.

Note that Eq.~(\ref{eq:heff}) is a special case of Eq.~(\ref{eq:hambipart}).
This suggests that starting from a general $H_\text{DH}$ of the form given by Eq.~(\ref{eq:hambipart}) and running the argument similar to that leading from Eq.~(\ref{eq:hpxpring}) to Eq.~(\ref{eq:heff}) in reverse results in a series of {\it coupled} Hamiltonians whose action on $\ket*{\widetilde\Lambda}_\mathcal{F}$ is identical to that of $H_\text{DH}$. Let us now formalize the process of going from a decoupled $H_\text{DH}$ to a coupled Hamiltonian sharing an eigenstate $\ket*{\widetilde\Lambda}_\mathcal{F}$ with it.

\section*{Geometric generalization}
\addcontentsline{toc}{section}{Geometric generalization}
Consider general PXP-type Hamiltonians of the form
\begin{equation}
  \label{eq:pxpgen}
  \widetilde H_\text{PXP}(G)=\sum_{i \in V}X_i\prod_{j \in V: (i,j)\in E} P_j,
\end{equation}
where $G=(V,E)$ is an undirected graph whose vertices $V$ represent spins, and edges $E$ represent unordered interactions --- Rydberg blockades --- between pairs of spins.

Per earlier arguments, $\ket*{\widetilde\Lambda}_{\mathcal{F}}$ in Eq.~(\ref{eq:lambdagen}) is a zero-energy eigenstate of the DH Hamiltonian $\widetilde H_\text{PXP}(G_\mathcal{A} \cup G_\dualmathcal{A})$, where $G_\mathcal{A} \cong G_\dualmathcal{A}$ are isomorphic graphs, $\cup$ denotes a union of all vertices and edges, and $\mathcal{F}$ is a bitstring set generating a Krylov subspace $\mathcal{K}_\mathcal{F} = \operatorname{span}\{\ket{f}:f\in\mathcal{F}\}$ of $\widetilde H_\text{PXP}(G_\mathcal{A})$.
Let the graph isomorphism be established by the correspondence between vertices with labels $i$ and $\dual i$, and let $\ket{f}_\mathcal{A}\otimes\ket{f}_\dualmathcal{A}$ terms in Eq.~(\ref{eq:lambdagen}) be of the form $\ket{s_{1} s_{2} \dots s_{L}}_{1,2,\dots,L}\otimes \ket{s_1 s_2 \dots s_L}_{\dual 1,\dual 2,\dots,\dual L}$.

Because of the perfect correlation of spins $i$ and $\dual i$, the action of $\widetilde H_\text{PXP}(G_\mathcal{A} \cup G_\dualmathcal{A})$ on $\ket*{\widetilde\Lambda}_\mathcal{F}$ is equivalent to that of $\widetilde H_\text{PXP}(G_\mathcal{C})$ for
\begin{equation}
  \label{eq:gc}
  G_\mathcal{C} = \left(\cdots \circ\mathcal{W}^{(a_2,b_2)}_{\nu_2}
  \circ \mathcal{W}^{(a_1,b_1)}_{\nu_1}\right)
  \left(G_\mathcal{A} \cup G_\dualmathcal{A}\right),
\end{equation}
where $\mathcal{W}^{(a,b)}_{\nu}$ is a ``graph modifying''  operator acting on graph $G=(V, E)$. Its action is nontrivial only if $(a, \dual b), (\dual a, b) \in E$ (assuming $\dual{\dual i} \equiv i$) and given by
\begin{subequations}
  \label{eq:gmo}
  \begin{align}
    \label{eq:gmo_plus}
    &\mathcal{W}_+^{(a,b)}(G) =
      (V, E \cup \{(a, b)\}),\\
   \label{eq:gmo_minus}
    &\mathcal{W}_-^{(a,b)}(G) = 
      (V, E \setminus \{(a, b)\}).
  \end{align}
\end{subequations}
The $\nu=\text{``$+$''}$ branch of  $\mathcal{W}^{(a,b)}_{\nu}$ (assuming vertices $a$ and $b$ belong to different initial isomorphic graphs $G_\mathcal{A}$ and $G_\dualmathcal{A}$) adds an intersystem interaction duplicating existing intrasystem interactions from the perspective of $\ket*{\widetilde\Lambda}_\mathcal{F}$ [i.e., decorates the $X_a$ and $X_b$ operators in Eq.~(\ref{eq:pxpgen}) with superfluous for $\ket*{\widetilde\Lambda}_\mathcal{F}$ projectors], whereas the $\nu=\text{``$-$''}$ branch (assuming both $a$ and $b$ belong to either $G_\mathcal{A}$ or $G_\dualmathcal{A}$) removes an existing intrasystem interaction given a pair of equivalent for $\ket*{\widetilde\Lambda}_\mathcal{F}$ intersystem interactions exists [see Sec.~\suppref{supp:wop} of \cite{supp} for intuitive visual explanation of $\mathcal{W}^{(a,b)}_{\nu}$].

\begin{figure}
  \subfloat{\label{fig:ring}}
  \subfloat{\label{fig:dangler}}
  \subfloat{\label{fig:grid2d}}
  \subfloat{\label{fig:cylinder}}\par\nointerlineskip
  \includegraphics[width=\columnwidth]{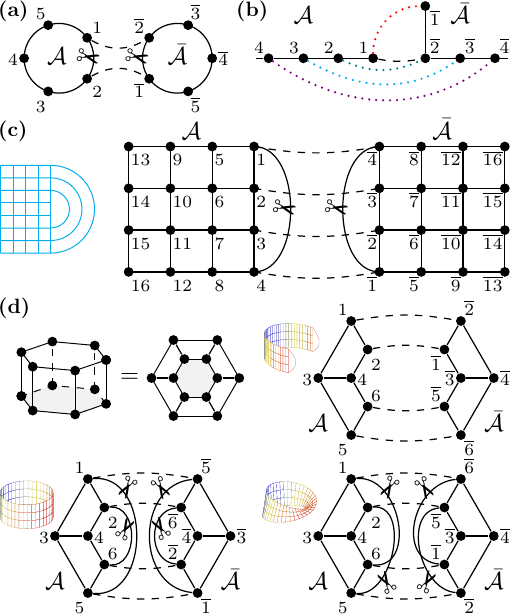}
  \caption{Constructions of systems with $\ket*{\widetilde\Lambda}_\mathcal{F}$ eigenstates.
  The edges of graphs $G_\mathcal{A}$ and $G_\dualmathcal{A}$ are solid lines; edges added via $\mathcal{W}_+^{(a,b)}$ are dashed lines; edges removed via $\mathcal{W}_-^{(a,b)}$ are indicated by the ``scissors'' symbol.
  (a)~PBC chain is produced by acting on $G_\mathcal{A} \cup G_\dualmathcal{A}$ with $\mathcal{W} = \mathcal{W}_-^{(\dual 1, \dual 2)} \circ  \mathcal{W}_-^{(1, 2)} \circ \mathcal{W}_+^{(\dual 1, 2)} \circ \mathcal{W}_+^{(1, \dual 2)}$.
  Clearly, $H_\text{PXP}\left[\mathcal{W}\left(G_\mathcal{A} \cup G_\dualmathcal{A}\right)\right]$ is identical to $H_\text{PXP}(N=10)$ given in Eq.~(\ref{eq:hpxpring}) up to the labels.
  (b)~The dangler is generated from an OBC chain (shown as system $\mathcal{A}$).
  The dotted lines indicate the pattern of perfectly correlated spins.
  (c)~$2L \times M$ square lattice. Although system $\mathcal{A}$ (whose geometry for a larger system is sketched on the left) is not itself a regular square lattice, its ``irregular'' interactions enable the construction.  
  When the height $M > 4$, such interactions are between sites $(1, M), (2, M-1), \dots, (M/2 - 1, M/2 + 2)$ on top of an otherwise regular square lattice.
  (d)~Distinct eigenstates of the PXP model on a $2L \times M$ cylinder, see text for details.
  }
  \label{fig:complexgeom}
\end{figure}

In Fig.~\ref{fig:ring}, we show how Eq.~(\ref{eq:gc}) leads from $G_\mathcal{A} \cup G_\dualmathcal{A}$ representing two decoupled PBC chains of size $L$ to $G_\mathcal{C}$ representing a PBC chain of size $2L$; by construction, both $\widetilde H_\text{PXP}(G_\mathcal{A} \cup G_\dualmathcal{A})$ and $\widetilde H_\text{PXP}(G_\mathcal{C})$ share an eigenstate $\ket*{\widetilde\Lambda}_{\mathcal{F}_L^{\text{PBC}}} \cong \ket{\Lambda}$ (as we already knew).
In general, for $\forall G_\mathcal{A}$, there exists a Krylov subspace $\mathcal{K}_\mathcal{F}$ of $\widetilde H_\text{PXP}(G_\mathcal{A})$ and, hence, $\ket*{\widetilde\Lambda}_\mathcal{F}$ is an eigenstate of $\widetilde H_\text{PXP}(G_\mathcal{A} \cup G_\dualmathcal{A})$ and all (exponentially many) coupled Hamiltonians that can be generated via Eq.~(\ref{eq:gc}).
For PXP-type models, the typical choice for $\mathcal{K}_\mathcal{F}$ is the natural generalized Rydberg-blockaded subspace, which we will denote by $\mathcal{R}$ for contextually implied systems.

A question of practical relevance is what physically reasonable systems besides the ring host eigenstates $\ket*{\widetilde\Lambda}_\mathcal{F}$.
Evidently, given any graph $G$ with arbitrary labeling of vertices, if there exists a graph $G_\mathcal{A}$ which generates a graph $G_\mathcal{C} \cong G$ via Eq.~(\ref{eq:gc}), then $\widetilde H_\text{PXP}(G)$ must host an eigenstate $\ket*{\widetilde\Lambda}_\mathcal{F}$ for any $\mathcal{K}_\mathcal{F}$ of $\widetilde H_\text{PXP}(G_\mathcal{A})$.
The spin pairing pattern of $\ket*{\widetilde\Lambda}_\mathcal{F}$ on $G$ is given by the bijection $(i, \dual i)_{G_\mathcal{C}} \leftrightarrow (a, b)_{G}$ relating labels of $G_\mathcal{C}$ and $G$.

In Sec.~\suppref{supp:noobc} of \cite{supp} we prove that no $G_\mathcal{A}$ generates an OBC chain; however, an irregular OBC chain hosting $\ket{\Lambda_\text{d}}\equiv\ket*{\widetilde\Lambda}_{\mathcal{F}_L^\text{OBC}}$, the ``dangler'' shown in Fig.~\ref{fig:dangler}, results from $G_\mathcal{A}$ that is itself a regular OBC chain \footnote{The set ${\mathcal{F}_L^\text{OBC}}: \mathcal{R} = \operatorname{span}\{\ket{f}:f\in\mathcal{F}_L^\text{OBC}\}$ is defined completely analogously to ${\mathcal{F}_L^\text{PBC}}$ for OBC chains of size $L$. Note that $\ket*{\widetilde\Lambda}_{\mathcal{F}_L^\text{OBC}}$ saturates the maximal entanglement entropy attainable for the standard bipartition. In Sec.~\suppref{supp:mps} of \cite{supp} we show that $\ket*{\widetilde\Lambda}_{\mathcal{F}_L^\text{OBC}}$ has an MPS form with the same bulk tensors as in the TI MPS representation of $\ket*{\widetilde\Lambda}_{\mathcal{F}_L^\text{PBC}}$.}.
Notably, this system breaks the inversion symmetry and hence its zero energy manifold is not exponentially large like in the case of the OBC chain~\cite{Turner_2018quantum,Buijsman_2022}; thus, exponential degeneracy of the nullspace is neither a prerequisite nor a guarantee for the existence of an eigenstate $\ket*{\widetilde\Lambda}_\mathcal{F}$.

Constructions in Figs.~\ref{fig:grid2d} and \ref{fig:cylinder} reveal $\ket*{\widetilde\Lambda}_\mathcal{F}$ eigenstates in the spectra of $\widetilde H_\text{PXP}(G)$ where $G$ is a square lattice with OBCs, and a cylinder (exemplified for simplicity by a PBC two-leg ladder). Generically, the choice of $G_\mathcal{A}$ generating a $G_\mathcal{C} \cong G$ is not unique. For instance, in Fig.~\ref{fig:grid2d} the cut could be horizontal instead of vertical, in which case $G_\mathcal{A}$ would be an $8 \times 2$ lattice with three ``irregular'' interactions.
However, such construction would yield an identical geometric pairing pattern of perfectly correlated spins on $G$ to the one demonstrated. In fact, we can prove that for any PXP system at most one $\ket*{\widetilde\Lambda}_\mathcal{F} \in \mathcal{R}$ can have a given pairing pattern (see Sec.~\suppref{supp:uniqueallzz} of~\cite{supp}).

Remarkably, the PBC two-leg ladder (with an even number of sites along its direction with PBCs) allows for multiple linearly independent $\ket*{\widetilde\Lambda}_\mathcal{F}$ eigenstates since the three choices of $G_\mathcal{A}$ (OBC, PBC, and M\"{o}bius strip ladders) lead to distinct nonoverlapping pairing patterns on $G$. 
The construction with $G_\mathcal{A}$ being an OBC ladder is close to that in Fig.~\ref{fig:grid2d} with the difference being that in the former case the ``stitching'' of $G_\mathcal{A}$ and $G_\dualmathcal{A}$ is performed across two boundaries; the construction with a PBC ladder is similar to that in Fig.~\ref{fig:ring}, whereas the construction with a M\"{o}bius strip is unique to this system. 
The PBC ladder and M\"{o}bius strip constructions yield distinct TI along the PBC direction eigenstates.
The OBC ladder construction results in $L$ linearly independent translational-symmetry-breaking $L$-periodic eigenstates.
Thus, we get the total of $L+2$ linearly independent eigenstates residing in $\mathcal{R}$.
The PBC and M\"{o}bius strip ladder constructions extend to the cylinder without any restriction on the height $M$, whereas the OBC ladder construction works only when $M$ is even.
In the latter case, $G_\mathcal{A}$ becomes an OBC lattice with ``irregular'' interactions like those in Fig.~\ref{fig:grid2d} at each boundary --- as shown in the corresponding three-dimensional drawing in Fig.~\ref{fig:cylinder}.

Although the three choices of $G_\mathcal{A}$ in Fig.~\ref{fig:cylinder} yield globally distinct pairing patterns, correlated spins $(3,\dual 3)$ and $(4,\dual 4)$ are shared by the OBC ladder or M\"{o}bius strip constructions.
This means that the simultaneous eigenspace of the PXP, $Z_{3}Z_{\dual 3}$, and $Z_{4}Z_{\dual 4}$ Hamiltonians is at least 2-dimensional \footnote{In general, when $L$ is odd and $M$ is even, the two constructions share $M$ correlated spins pairs; e.g., $(L,\dual L)$ and $(L+1,\dual {L+1})$ when $M=2$ and the site labeling is like in Fig.~\ref{fig:cylinder}.}.
In fact, in all numerically accessible systems we probed, the converse statement is also true: given any graph $G$ and its distinct vertices $a$ and $b$, the simultaneous eigenspace of $\widetilde H_\text{PXP}(G)$ and $Z_{a}Z_{b}$ on $\mathcal{R}$ is spanned by $\ket*{\widetilde\Lambda}_\mathcal{F}$ states with perfectly correlated spins $a$ and $b$.
While we cannot prove that this is always true, the following theorem (proven in Sec.~\suppref{supp:dangler} of~\cite{supp}) addresses one special case.
\begin{theorem}
  \label{thrm:dangler}
  Let $G_\mathcal{C}$ be the dangler graph [as in Fig.~\ref{fig:dangler}]. The simultaneous eigenspace of $H_\mathrm{d}\equiv\widetilde H_\mathrm{PXP}(G_\mathcal{C})$ and $Z_1Z_{\dual 1}$ on $\mathcal{R}$ is $\operatorname{span}\{\ket{\Lambda_\mathrm{d}}\}$.
\end{theorem}

We emphasize that Theorem~\ref{thrm:dangler} is a narrow special case of a frequent pattern (e.g., we find numerically that the same holds for the PBC chain). 
It is provided to demonstrate that such proofs can, in principle, be constructed, and to motivate the following discussion.

\section*{Experimental relevance}
\begin{figure}
  \subfloat{\label{fig:prep}}
  \subfloat{\label{fig:quench}}\par\nointerlineskip
  \includegraphics[width=\columnwidth]{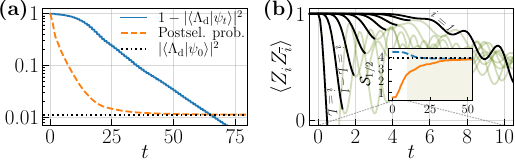}
  \caption{
    Numerical study of an $N=18$ dangler system.
    (a)~Starting from $\ket{\psi_0} = \ket{00\ldots 0}$, $\ket{\Lambda_\text{d}}$ state is prepared via postselection on the $+1$ outcomes of $Z_1Z_{\dual 1}$ measurements preceded by evolution under the native PXP Hamiltonian over period $\tau=1$.
    The weight of the component of the wave function orthogonal to $\ket{\Lambda_\text{d}}$ (infidelity) becomes exponentially small in the number of successful measurements, while the overall postselection probability converges to $|\braket{\Lambda_\text{d}}{\psi_0}|^2$.
    (b)~Quench from the initial state $Z_L\ket{\Lambda_\text{d}}$ (here $L = 9$).
    The curves (partially in muted tones for clarity) show sequential decay of spin-spin correlations starting from $\ev{Z_LZ_{\dual L}}$ and up to $\ev{Z_1Z_{\dual 1}}$ as the ``butterfly'' cone spreads through the system.
    The inset shows the half-system entanglement entropies for the standard (dashed line) and entropy-minimizing (solid line) bipartitions; information scrambling makes both entropies approach the Page value \cite{Page_1993}, $\mathcal{S}_P \simeq \log\left\vert\mathcal{F}^\text{OBC}_L\right\vert - 1/2$ (dotted line), expected for bipartitions of random featureless states.
  }
\end{figure}

\addcontentsline{toc}{section}{Experimental relevance}
With the dangler system in mind, consider non-Hermitian operator $M_\tau = \mathcal{P}^+_{1\dual 1} e^{-iH_\text{d}\tau}$ that applies projector $\mathcal{P}^+_{1\dual 1} = \ketbra{00}{00}_{1\dual 1} + \ketbra{11}{11}_{1\dual 1}$ after time evolution under $H_\text{d}$.
In Sec.~\suppref{supp:mtau} of \cite{supp} we argue that per Theorem~\ref{thrm:dangler}, for generic $\tau > 0$, $\lim_{k\to\infty} M_\tau^k = \ketbra{\Lambda_\text{d}}{\Lambda_\text{d}}$; i.e., postselection on the $+1$ outcomes of a sequence of $Z_1 Z_{\dual 1}$ measurements preceded by evolution under $H_\text{d}$ over period $\tau$ converges to the projector onto the $\ket{\Lambda_\text{d}}$ eigenstate [see Fig.~\ref{fig:prep}].
Thus, the maximally entangled state $\ket{\Lambda_\text{d}}$ can be prepared via strictly local measurements of only two spins with fidelity limited only by the experimental imperfections.
A similar protocol for preparing a rainbow eigenstate of the unconstrained 2D XY Hamiltonian was also proposed in \cite{agarwal2023longrange}.

A system prepared in an eigenstate $\ket*{\widetilde\Lambda}_\mathcal{F}$ enables access to richer dynamics in comparison to the one prepared in a simple product state.
For example, suppose the initial state $\ket{\Lambda_\text{d}}$ is suddenly changed into another maximally entangled state $Z_L\ket{\Lambda_\text{d}}$ that is no longer an eigenstate of $H_\text{d}$.
The effect of the perturbing operator $Z_L$ acting on the leftmost site of the system [see Fig.~\ref{fig:dangler}] is indistinguishable from that of $Z_{\dual L}$ acting on the opposite side.
This implies that the light cone spreads from both edge sites $L$ and $\dual L$ towards the center of the system sequentially destroying perfect correlations between spin pairs $(i,\dual i)$ [Fig.~\ref{fig:quench}]\nocite{Page_1993}.
In Sec.~\suppref{supp:otoc} of \cite{supp} we argue that over some initial time interval the cross-subsystem correlation functions $\ev{Z_iZ_{\dual i}}$ plotted in Fig.~\ref{fig:quench} are directly related to the out-of-time-order correlator (OTOC)~\cite{xu2024scrambling} $\ev{[Z_i(t), Z_L(0)]^2}, i \in [1,\dots,L]$, for the infinite-temperature Gibbs ensemble of an OBC chain of size $L$ (such as the subsystem $\mathcal{A}$); we also discuss how the OTOCs can be measured exactly using the corresponding $H_{\text{DH}}$. Note that our OTOC measurement protocol does not require backward time evolution (cf.~Refs.~\cite{Sundar_2022,green2022experimental}; see Refs.~\cite{swingle2016measuring,yao2016interferometric,dag2019detection,yoshida2019disentangling,yuan2022quantum} for other related OTOC topics).
In Secs.~\suppref{supp:stability} and~\suppref{supp:exp} of \cite{supp} we numerically study the effect of perturbations on the state preparation and quench protocols and also discuss several practical enhancements.

We conclude by noting that while we focused on the state preparation and subsequent dynamics experiments in the specific dangler geometry, similar physics holds under appropriate conditions in other systems hosting $\ket*{\widetilde\Lambda}_\mathcal{F}$ scars discussed in this Letter.

\section*{Acknowledgments}
We thank Sanjay Moudgalya, Cheng-Ju Lin, Manuel Endres, Daniel Mark,  Federica Surace, Pablo Sala, Sara Vanovac, and Leo Zhou for useful discussions and previous collaborations on related topics.
This work was supported by the National Science Foundation through Grant No. DMR-2001186.

\bibliography{main}
\ifnosupp
  \dummylabel{supp:proofalt}{I}
  \dummylabel{supp:entstr}{II}
  \dummylabel{supp:mps}{III}
  \dummylabel{supp:wop}{IV}
  \dummylabel{supp:noobc}{V}
  \dummylabel{supp:uniqueallzz}{VI}
  \dummylabel{supp:dangler}{VII}
  \dummylabel{supp:mtau}{IX}
  \dummylabel{supp:otoc}{X}
  \dummylabel{supp:stability}{XI}
  \dummylabel{supp:exp}{XII}
\else
  \input{supp.tex}
\fi

\end{document}

%% file: supp.tex
\clearpage
\title{\large\bfseries Supplemental Material: Volume-entangled exact scar states in the PXP and related models in any dimension}
\maketitle

\titleformat{\section}[block]{\normalfont\bfseries\filcenter}{\thesection.}{1em}{\MakeUppercase}
\titleformat{\subsection}[block]{\normalfont\bfseries\filcenter}{\thesubsection.}{1em}{}

\titlespacing*{\section}
              {0em}
              {2em}
              {1em}

\renewcommand{\thepage}{S\arabic{page}}
\renewcommand{\thesection}{\Roman{section}}
\renewcommand{\thefigure}{S\arabic{figure}}
\renewcommand{\theHfigure}{S\arabic{figure}}
\renewcommand{\theequation}{S\arabic{equation}}

\setcounter{section}{0}
\setcounter{equation}{0}
\setcounter{figure}{0}
\setcounter{page}{1}

\setcounter{secnumdepth}{1}

\onecolumngrid

\section{Alternative proof that \texorpdfstring{$\ket{\Lambda}$}{state Lambda} is an eigenstate of \texorpdfstring{$H_\text{PXP}(N)$}{PXP Hamiltonian}}
\label{supp:proofalt}
In this section we provide an alternative proof of a slightly more general version of Theorem~\suppref{thrm:lambda} in the main text.
In particular, we show that $\ket{\Lambda}$ is a zero energy eigenstate of $P_{i-1} X_i P_{i+1} + P_{L+i-1} X_{L+i} P_{L+i+1}$ for $\forall i \in \{1,\dots, L\}$, which means that terms in the full PXP Hamiltonian are required to be identical only for sites $i$ and $L+i$.

To simplify the notation and without loss of generality (since $\ket{\Lambda}$ is translationally invariant), we can take $i=1$. Let us express $\ket{\Lambda}$ as
\begin{equation}
  \label{eq:lambda11}
  \ket{\Lambda} =
  \frac{1}{\sqrt{\chi}}\sum_{f \in
    \mathcal{F}^{\text{PBC}}_L}(-1)^{|f|}\ket{f_1f_1}_{1,L+1}\otimes\ket{f_2\dots f_L}_{2,\ldots,L}\otimes\ket{f_2\dots f_L}_{L+2,\ldots,2L},
\end{equation}
where $f_j$ is the $j^\text{th}$ bit in $f$.
We can reorganize the summation in Eq.~(\ref{eq:lambda11}) by using bitstrings $f$ with $f_1 = 0$ as follows:
\begin{equation}
  \label{eq:lambdaf1}
  \ket{\Lambda} =
  \frac{1}{\sqrt{\chi}}\sum_{f \in
    \mathcal{F}^{\text{PBC}}_L: f_1 = 0}(-1)^{|f|}
  \left\{\begin{array}{ll}    
    \ket{00}-\ket{11}, & \text{if } f_2=f_L=0, \\
    \ket{00}, & \text{otherwise}
  \end{array}\right\}_{1,L+1}
  \otimes\ket{f_2\dots f_L}_{2,\ldots,L}\otimes\ket{f_2\dots f_L}_{L+2,\ldots,2L}.
\end{equation}
In the terms corresponding to the first case above, the spin at site $1$ ($L+1$) is not Rydberg-blockaded by the spins at sites $2L$ and $2$ ($L$ and $L+2$), which means $\ket{11}_{1,L+1}\otimes\ket{f_2\dots f_L}_{2,\ldots,L}\otimes\ket{f_2\dots f_L}_{L+2,\ldots,2L}$ with the opposite sign must be included in the summation as shown.

Clearly, all individual terms in the summation in Eq.~(\ref{eq:lambdaf1}) are annihilated by $P_{2L} X_1 P_2 + P_L X_{L+1} P_{L+2}$.
In the first case, due to the absence of excitations adjacent to sites $1$ and $L+1$, $P_{2L} X_1 P_2 + P_L X_{L+1} P_{L+2}$ acts as $X_1 + X_{L+1}$, which results in total destructive interference (when collecting contributions to amplitudes of states produced by the action of the individual terms).
In the second case, excitations adjacent to sites $1$ and $L+1$ prevent both $P_{2L} X_1 P_2$ and $P_L X_{L+1} P_{L+2}$ from producing any new states.

The above argument generalizes to models with extended Rydberg blockades in arbitrary dimension.
In particular, per Lemma~\ref{lemma:localpxp} in Sec.~\ref{supp:uniqueallzz} of this Supplemental Material, eigenstates of the form given in Eq.~(\suppref{eq:lambdagen}) of the generalized PXP Hamiltonian in Eq.~(\suppref{eq:pxpgen}) are also eigenstates of a related family of Hamiltonians with arbitrary coefficients that match for terms corresponding to pairs of correlated spins.

\section{Entanglement structure of \texorpdfstring{$\ket{\Lambda}$}{state Lambda}}
\label{supp:entstr}
\begin{figure}
  \includegraphics[width=0.45\textwidth]{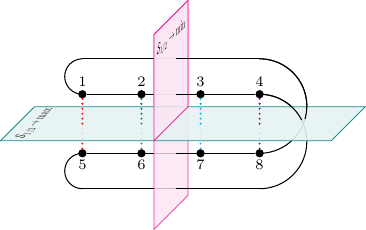}
  \caption{
    Standard (horizontal plane) and entanglement-minimizing (vertical plane) bipartitions of $\ket{\Lambda}$. Note that this is the same chain as in Fig.~\suppref{fig:pbcchain} of the main text, deformed for demonstration purposes. The entanglement entropy of the standard bipartition is linear in the system size (i.e., as volume law), whereas the fine-tuned entanglement-minimizing cut gives area-law scaling. The entanglement-minimizing bipartition does not split any perfectly correlated spin pairs and for that reason the entanglement entropy associated with it is purely due to the kinetic constraint (Rydberg blockade) of the Hilbert space.
  }
  \label{fig:bipart}
\end{figure}

In this section we derive various results characterizing entanglement in the state $\ket{\Lambda}$ on the PBC chain of size $N=2L$.
One of the prominent features designating $\ket{\Lambda}$ as an ETH-violating scar state of the PXP model is its cut-dependent entanglement entropy anomalous for eigenstates in the middle of the energy spectrum. As we show in Fig.~\ref{fig:bipart}, the \emph{standard} (entanglement-maximizing) and fine-tuned \emph{entanglement-minimizing} bipartitions of this system into two equal halves exhibit, respectively, volume- and area-law scaling of entanglement entropy.
The standard bipartition cuts two physical bonds and maximizes the entanglement by splitting each perfectly correlated spin pair across the two subsystems, whereas the entanglement-minimizing bipartition achieves the exact opposite (i.e., confines every perfectly correlated spin pair to one of the subsystems) by cutting four fine-tuned physical bonds.

From the form of Eq.~(\suppref{eq:lambda}) we immediately see that the Schmidt values for the standard bipartition are all equal to $\chi^{-1/2}$, where $\chi = F_{L-1}+F_{L+1}$ is the Schmidt index.
Therefore, the entanglement entropy of the standard bipartition is $\mathcal{S}_{1/2}^\text{standard} = \log\chi$, which is linear in the system size for large $N$.
Let us put this in perspective.
The entanglement entropy of the standard bipartition of any state defined on a PBC chain subject to Rydberg blockade into two half-systems is upper-bounded by $\mathcal{S}_{\max} = \log\left|\mathcal{F}^{\text{OBC}}_L\right|$, where $\left|\mathcal{F}^{\text{OBC}}_L\right| = F_{L+2}$ is the cardinality of the set $\mathcal{F}^{\text{OBC}}_L$ containing all bitstrings defining the Rydberg-blockaded subspace $\mathcal{R}$ for systems of $L$ spins with OBCs, i.e., the maximum possible dimension of the Hilbert space of a connected half-system.
In Ref.~\suppcite{Page_1993} it was argued that the average entanglement entropy of any bipartition, the so-called Page value, for a random pure state is given by
\begin{equation}
  \label{eq:page}
  \mathcal{S}_P \simeq \ln \mathcal{D}_\mathcal{A} - \frac{\mathcal{D}_\mathcal{A}}{2\mathcal{D}_\mathcal{B}},
\end{equation}
where $\mathcal{D}_\mathcal{A}$ and $\mathcal{D}_\mathcal{B}$ are the dimensions of the Hilbert spaces of the two subsystems.
Assuming $\mathcal{D}_\mathcal{A} = \mathcal{D}_\mathcal{B} \simeq \left|\mathcal{F}^{\text{OBC}}_L\right|$, and disregarding the fact that the derivation of Eq.~(\ref{eq:page}) did not account for constrained Hilbert spaces without tensor product structure, we can take $\mathcal{S}_P \simeq \mathcal{S}_{\max} - \frac{1}{2}$.
It easy to show using Binet's formula $F_n=(\varphi^n-(-\varphi)^{-n})/\sqrt{5}$, where $\varphi = (1 + \sqrt{5})/2$ is the golden ratio, that $\mathcal{S}_{\max} - \mathcal{S}_{1/2}^\text{standard} \to -\log(\varphi^{-3}+\varphi^{-1})\approx 0.16$ for large $N$.
Thus, $\mathcal{S}_{1/2}^\text{standard}$ has the highest rate of growth with systems size attainable on $\mathcal{R}$, and $\mathcal{S}_P < \mathcal{S}_{1/2}^\text{standard} < \mathcal{S}_{\max}$ (by small constants negligible in the thermodynamic limit).
We can, therefore, say that with respect to the standard bipartition into two half-systems (and, as will be shown later, with respect to any standard bipartition), $\ket{\Lambda}$ resembles a random maximally entangled pure state.

The exact entropy of the entanglement-minimizing bipartition into two half-systems --- which, due to the Rydberg blockade, we expect to be non-zero, in contrast with the rainbow state from Ref.~\suppcite{Langlett_2022} --- is somewhat harder to compute.
In Sec.~\ref{supp:mps} of this Supplemental Material, we obtain an MPS representation of $\ket{\Lambda}$ and use it to compute the exact value of this entropy, which, in the thermodynamic limit, is found to saturate at $\mathcal{S}_{1/2}^\text{ent.-min.} \approx 0.5736$ for even $L$, and at $\mathcal{S}_{1/2}^\text{ent.-min.} \approx 0.8763$ for odd $L$, consistent with area-law scaling.
For comparision, the exact scar state $\ket{\Phi_1}$ of the PXP chain with PBCs found in Ref.~\suppcite{Lin_2019} has the entropy of entanglement corresponding to its standard bipartition (entanglement-minimizing for this state) equal to $2\log 2\approx 1.3863$ (for even $L$).
Considering this, we can argue that $\ket{\Lambda}$ is, in fact, the least entangled and least thermal (and, hence, the ``simplest'') highly excited eigenstate of the PXP chain that conspires to appear maximally entangled and completely thermal to the local observer.
For a more detailed discussion of entanglement-minimizing bipartitions of $\ket{\Lambda}$ see Sec.~\ref{supp:mps} of this Supplemental Material.

In the remainder of this section we will derive several additional results further characterizing the state $\ket{\Lambda}$.
\subsection{Reduced density matrices (RDMs) for standard bipartitions}
The partial trace of the form $\rho_m =\Tr_B\ketbra{\Lambda}{\Lambda}$, where subsystem $B$ is a geometrically connected subset of $N - m \geq N/2$ sites (i.e., the standard bipartition where the subsystems are not necessarily of the same size) is diagonal; by definition
\begin{equation}
  \label{eq:rdm}
  \rho_m = \sum_{\substack{f \in
      \mathcal{F}^{\text{PBC}}_L,\\g=f_{[m+1\dots
        L]}}}\left(\mathbf{1}_m\otimes\bra{gf}\right)\ketbra{\Lambda}{\Lambda}
  \left(\mathbf{1}_m\otimes\ket{gf}\right)  =
  \frac{1}{\chi}\sum_{\substack{f \in
      \mathcal{F}^{\text{PBC}}_L,\\h=f_{[1\dots m]}}}\ketbra{h}{h} =
  \frac{1}{\chi}\sum_{h \in
    \mathcal{F}^{\text{OBC}}_m}\alpha_m(h)\ketbra{h}{h},
\end{equation}
where
\begin{equation}
  \label{eq:alpha}
  \alpha_m(h) = F_{L - m + 2 - (h_1 + h_m)}
\end{equation}
and subscripts attached to bitstrings extract a subset of bits with the specified indices; specifically, in Eq.~(\ref{eq:alpha}), $h_1$ and $h_m$ are the values of the first and $m^\text{th}$ bits in $h$ interpreted as integers.
In particular,
\begin{equation}
  \rho_1 =
  \frac{1}{\chi}\left(F_{L+1}\ketbra{0}{0}+F_{L-1}\ketbra{1}{1}\right),
\end{equation}
which in the thermodynamic limit yields the following RDM corresponding to an infinite-temperature Gibbs ensemble:
\begin{equation}
  \label{eq:rho1inf}
  \rho_1^\infty =
  \frac{1}{\sqrt{5}}(\varphi\ketbra{0}{0}+\varphi^{-1}\ketbra{1}{1}).
\end{equation}
\subsection{Scaling of the von Neumann entropy with subsystem size for standard bipartitions}
From Eqs.~(\ref{eq:rdm}) and (\ref{eq:alpha}) we see that $\rho_m$ will have $F_{m+2}$ diagonal entries (corresponding to the cardinality of $\mathcal{F}^{\text{OBC}}_m$), of which $F_m$ entries with $h_1+h_m=0$ will have $\alpha_m(h) = F_{L-m+2}$; $2F_{m-1}$ entries with $h_1+h_m=1$ will have $\alpha_m(h) = F_{L-m+1}$; and $F_{m-2}$ entries with $h_1+h_m=2$ will have $\alpha_m(h) = F_{L-m}$.
Thus the Von Neumann entropy for the standard bipartition into subsystems of sizes $1\leq m \leq L$ and $N-m$ is
\begin{equation}
  \label{eq:sent0}
  \begin{aligned}
    \mathcal{S}(\rho_m) &= -\Tr[\rho_m\log\rho_m]
    \\ &=-\frac{1}{\chi}\left(
         \begin{aligned}[t]
           F_mF_{L-m+2}\log\frac{F_{L-m+2}}{\chi}
           +2F_{m-1}F_{L-m+1}\log\frac{F_{L-m+1}}{\chi}
           +F_{m-2}F_{L-m}\log\frac{F_{L-m}}{\chi}
         \end{aligned}\right).
  \end{aligned}
\end{equation}
For the last equation to be valid for $m=1$ and $m=L$, we need to assume that $F_{-1} = 1$ (true for the extension of the Fibonacci numbers to negative integers given by $F_{-n}=(-1)^{n+1}F_n$), and that $0\cdot\log 0 = 0$.
It follows from the discussion above that $F_mF_{L-m+2}+2F_{m-1}F_{L-m+1}+F_{m-2}F_{L-m} = \chi\cdot Tr\rho_m = \chi$ and, therefore, Eq.~(\ref{eq:sent0}) can also be written in a form that is more suitable for the case when $m$ is close to $L$ as follows:
\begin{equation}
  \label{eq:sent1}
  \mathcal{S}(\rho_m) = \log\chi
  -\frac{1}{\chi}\left(
    F_mF_{L-m+2}\log F_{L-m+2} +2F_{m-1}F_{L-m+1}\log
    F_{L-m+1} +F_{m-2}F_{L-m}\log F_{L-m} \right).
\end{equation}
In the thermodynamic limit, for any finite $m$, it is possible to show using exact expressions for $F_m, F_{m-1}, F_{m-2}$ and asymptotic expressions for $F_{L-m+2}, F_{L-m+1},F_{L-m}$ that are exponentially accurate in $L$ that Eq.~(\ref{eq:sent0}) reduces to
\begin{equation}
  \mathcal{S}(\rho_m) = \left(m - \frac{2\varphi}{\sqrt{5}}\right)\log\varphi + \frac{1}{2}\log 5,
\end{equation}
indicating that, as expected, entanglement entropy of $\ket{\Lambda}$ for standard bipartitions grows according to the volume law.
For $m \gg 1$, $\mathcal{S}(\rho_m)$ approaches (with a tiny, negligible in the thermodynamic limit, constant deviation) the corresponding Page value given by Eq.~(\ref{eq:page}).

\subsection{Special non-local bipartitions}
Of particular interest are bipartitions where one of the subsystems consists of two spins separated by some distance $\ell$, which we can take to be spins $1$ and $1+\ell$ w.l.o.g.\ and write the corresponding RDM as $\rho[1; 1+\ell]$.
Note that $\ket{\Lambda}$ is an eigenstate of the two-qubit SWAP gate acting on any two spins separated by distance $L$, which means that $\rho[1; 1+\ell] = \rho[1; 1+\ell \pm L]$.
Therefore, for any $\ell \neq L$ we can consider the two spins to be from the same contiguous half-system, and it follows from Eq.~(\ref{eq:rdm}) that
\begin{equation}
\rho[1; 1+\ell] = \rho[1; 1+(\ell \mod L)] = \sum_{s,s' \in \{0,1\}} \beta_\ell(s,s') \ketbra{ss'}{ss'}, \qquad \ell \neq L,
\end{equation}
where coefficients $\beta_\ell(s,s')$ result from taking a partial trace over a subsystem of size $L-2$, which excludes the specific spins at positions 1 and $l+\ell$, of an infinite-temperature Gibbs ensemble of size $L$.
Thus for all pairs of spins in $\ket{\Lambda}$, except those separated by distance $l=L$, RDMs are diagonal mixtures of product states.

For spins separated by distance $l=L$ the situation is very different.
To simplify the analysis, let us reorder the spins in $\ket{\Lambda}$ as follows: $\ket{s_1s_2\dots s_L}^{\otimes 2} \to \ket{s_1s_1}\otimes\ket{s_2\dots s_L}^{\otimes 2}$.
Then in this reordered basis $\ket{\Lambda}$ can be written as
\begin{equation}
  \ket{\Lambda} = \sum_{g \in \mathcal{F}_{L-1}^{\text{OBC}}} (-1)^{|g|}\left(\ket{00}-\delta_{g_1,0} \delta_{g_{L-1},0}\ket{11}\right)\otimes\ket{g}^{\otimes 2}.
\end{equation}
Taking a partial trace over the last $N-2$ spins, one gets the following RDM for sites $1$ and $L$:
\begin{equation}
  \label{eq:rho1L}
  \rho[1;1+L] = \frac{1}{\chi}\big[F_{L-1}(\ket{00} - \ket{11})(\bra{00} - \bra{11}) 
  + F_L \ketbra{00}{00}\big],
\end{equation}
which for sufficiently large $L$ is very close to
\begin{equation}
  \label{eq:rho1Linf}
  \lim_{L \to \infty} \rho[1;1+L] =\frac{1}{\sqrt{5}\varphi}\left(\varphi^2\ketbra{00}{00} + \ketbra{11}{11} - \ketbra{00}{11} - \ketbra{11}{00}\right).
\end{equation}
Note that reduction from Eq.~(\ref{eq:rho1Linf}) to single-spin RDM is in agreement with Eq.~(\ref{eq:rho1inf}).

In contrast with the rainbow scars constructed in~\suppcite{Langlett_2022}, the pairs of perfectly correlated spins in $\ket{\Lambda}$ are not maximally entangled Bell pairs.
Specifically, $\lim_{L \to \infty} \rho[1;L+1]$ is a mixture of two less-than-maximally entangled states, which up to normalization can be written as
\begin{equation}
  \label{eq:rho1Lpure}
  \begin{aligned}
    &\ket{\xi_1} = 2\ket{00} - \eta\ket{11}, \\
    &\ket{\xi_2} = \eta\ket{00} + 2\ket{11},
  \end{aligned}
\end{equation}
where $\eta = \sqrt{\varphi^4 - 2\varphi^2 + 5} - \varphi \approx 0.9545$;
the eigenvalue associated with $\ket{\xi_1}$ in $\lim_{L \to \infty} \rho[1;L+1]$ is $p_1=(2\varphi^2 + \eta)/(2\sqrt{5}\varphi)\approx 0.856$.
The discrepancy between the properties of the perfectly correlated spin pairs in $\ket{\Lambda}$ and in the rainbow state is entirely due to the fact that we restricted $\ket{\Lambda}$ to $\mathcal{R}$, where the dynamical constraint makes each individual spin (and, by extension, any subsystem) of PXP model's eigenstates entangled with the rest of the system (this is the reason the entanglement entropy of entanglement-minimizing bipartitions discussed earlier is non-zero).
Such a restriction, obviously inessential to the general construction, was chosen due to its relevance for experiments with Rydberg atom arrays.
By substituting $\mathcal{F}_L^{\text{PBC}}$ with $\{0, 1\}^{L}$ in Eq.~(\suppref{eq:lambda}) we get a zero energy eigenstate of $H_\mathrm{PXP}(N)$ in the space free from any dynamical constraint.
This eigenstate would be more rainbow-like; in particular, repeating the above analysis, one can easily show that $\rho[1;L+1]$ will change from the mixed state given in Eqs.~(\ref{eq:rho1L})--(\ref{eq:rho1Lpure}) to a pure Bell state proportional to $\ket{00}-\ket{11}$.

We conclude this section by noting that higher-dimensional generalizations of the state $\ket{\Lambda}$ have qualitatively similar entanglement structure to that described above.
In particular, they all weakly violate the ETH by virtue of having fine-tuned entanglement-minimizing bipartitions with area-law scaling of entanglement entropy. 

\section{Exact MPS representations of \texorpdfstring{$\ket{\Lambda} \cong \ket*{\widetilde\Lambda}_{\mathcal{F}_L^{\text{PBC}}}$}{state Lambda} and \texorpdfstring{$\ket{\Lambda_\text{d}} \equiv \ket*{\widetilde\Lambda}_{\mathcal{F}_L^{\text{OBC}}}$}{state Lambda d}}
\label{supp:mps}
Just like the area-law scar states introduced in Ref. \suppcite{Lin_2019}, $\ket{\Lambda}$ on the PBC chain [depicted  in Fig.~\suppref{fig:pbcchain}] can be written as an exact TI MPS upon appropriate grouping of sites.
In this section, we will obtain MPS representations of two related states, which --- using the generalized Eq.~(\suppref{eq:lambdagen}) and assuming perfectly correlated sites in subsystems $\mathcal{A}$ and $\dualmathcal{A}$, respectively, are labeled as $i$ and $\dual i$ --- we write as $\ket*{\widetilde\Lambda}_{\mathcal{F}_L^{\text{PBC}}}$ and $\ket*{\widetilde\Lambda}_{\mathcal{F}_L^{\text{OBC}}}$.
The former is identical to $\ket{\Lambda}$ as defined in Eq.~(\suppref{eq:lambda}) and Fig.~\suppref{fig:pbcchain} of the main text (and also in Secs.~\ref{supp:proofalt} and \ref{supp:entstr} of this Supplemental Material) under the mapping $\dual i \to i + L$ [cf. Fig.~\suppref{fig:ring}]; the latter can be interpreted as an OBC variant of $\ket*{\widetilde\Lambda}_{\mathcal{F}_L^{\text{PBC}}}$, or the eigenstate $\ket{\Lambda_\text{d}}$ of the so-called dangler system discussed in the main text and shown in Fig.~\suppref{fig:dangler}. We will also use the MPS form to obtain exact entanglement entropies corresponding to the entanglement-minimizing bipartitions of both states.

\begin{theorem}
In the basis of two-spin product states $\{\ket{st}: s,t \in \{0, 1\}\}$, site-independent matrices $M^{st}$ defined as
\begin{equation}
  \begin{aligned}
    M^{00} = \frac{1}{2}\begin{pmatrix}
      1 & 1 \\
      1 & 1
    \end{pmatrix},\quad
    M^{11} = -2\begin{pmatrix}
      0 & 0 \\
      1 & 0
    \end{pmatrix},\quad
    M^{01} = M^{10} = \mathbf{0}
  \end{aligned}
\end{equation}
generate $\ket{\Lambda}$ (up to normalization, and for $L > 1$) as follows:
\begin{equation}
  \label{eq:lambdamps}
  \ket{\Lambda} = \sum_{\{s_i\},\{s_{\dual i}\}}\Tr{M^{s_1s_{\dual 1}}M^{s_2s_{\dual 2}}\cdots M^{s_Ls_{\dual L}}}\ket{s_1s_2\dots s_L}\otimes \ket{s_{\dual 1}s_{\dual 2}\dots s_{\dual L}}.
\end{equation}
With terminations (boundary vectors) $v_\ell^T = v_r^T = (1/\sqrt{2}, 1/\sqrt{2})$, the state $\ket{\Lambda_\mathrm{d}}$ is given by
\begin{equation}
  \label{eq:lambdadmps}
  \ket{\Lambda_\mathrm{d}} = \sum_{\{s_i\},\{s_{\dual i}\}} v_\ell^T [M^{s_1s_{\dual 1}}M^{s_2s_{\dual 2}}\cdots M^{s_Ls_{\dual L}}] v_r \ket{s_1s_2\dots s_L}\otimes \ket{s_{\dual 1}s_{\dual 2}\dots s_{\dual L}}.
\end{equation}

\begin{proof}
  Note that matrices $M^{00}$ and $M^{11}$ are, respectively, idempotent and nilpotent: $\left(M^{00}\right)^k = M^{00}$ for $\forall k > 0$, whereas $\left(M^{11}\right)^2=\mathbf{0}$.
  This means that
  \begin{equation}
    \label{eq:lambdatr}
    \begin{aligned}
      \Tr{M^{s_1s_{\dual 1}}M^{s_2s_{\dual 2}}\cdots M^{s_Ls_{\dual L}}} &= \mathbf{1}_{\mathcal{F}^{\text{PBC}}_L}(s_1s_2\dots s_L)\cdot \left(\prod_{i=1}^L \delta_{s_i,s_{\dual i}}\right) \Tr{M^{00}\left(M^{00}M^{11}\right)^{\sum_{i=1}^L s_i}} \\
      &= \mathbf{1}_{\mathcal{F}^{\text{PBC}}_L}(s_1s_2\dots s_L)\cdot\left(\prod_{i=1}^L \delta_{s_i,s_{\dual i}}\right)\cdot(-1)^{\sum_{i=1}^L s_i},
    \end{aligned}
  \end{equation}
  where $\mathbf{1}_{\mathcal{F}^{\text{PBC}}_L}(s_1s_2\dots s_L) = [s_1s_2\dots s_L \in \mathcal{F}^{\text{PBC}}_L]$ is the indicator function for the set $\mathcal{F}^{\text{PBC}}_L$.
  Combining Eqs.~(\ref{eq:lambdamps}) and (\ref{eq:lambdatr}), we obtain
  \begin{equation}
    \label{eq:lambdampsfinal}
    \begin{aligned}
      \ket{\Lambda} &= \sum_{\{s_i\},\{s_{\dual i}\}}\mathbf{1}_{\mathcal{F}^{\text{PBC}}_L}(s_1s_2\dots s_L)\cdot\left(\prod_{i=1}^L \delta_{s_i,s_{\dual i}}\right)\cdot(-1)^{\sum_{i=1}^L s_i}\ket{s_1s_2\dots s_L}\otimes \ket{s_{\dual 1}s_{\dual 2}\dots s_{\dual L}}\\
      &= \sum_{\{s_i\}}\mathbf{1}_{\mathcal{F}^{\text{PBC}}_L}(s_1s_2\dots s_L)\cdot(-1)^{\sum_{i=1}^L s_i}\ket{s_1s_2\dots s_L}\otimes \ket{s_{1}s_{2}\dots s_{L}},
    \end{aligned}
  \end{equation}
  which, up to normalization, gives the same state as Eq.~(\suppref{eq:lambda}) in the main text.
  Following very similar steps, and using $v_\ell^TM^{00} = v_\ell^T$ and $M^{00} v_r = v_r$, one can show that Eq.~(\ref{eq:lambdadmps}) is equivalent to
  \begin{equation}
    \label{eq:lambdadmpsfinal}
      \ket{\Lambda_\text{d}} = \sum_{\{s_i\}}\mathbf{1}_{\mathcal{F}^{\text{OBC}}_L}(s_1s_2\dots s_L)\cdot(-1)^{\sum_{i=1}^L s_i}\ket{s_1s_2\dots s_L}\otimes \ket{s_{1}s_{2}\dots s_{L}},
  \end{equation}
  which agrees with our definition of $\ket{\Lambda_\text{d}}$.
\end{proof}
\end{theorem}
The $\ket*{\widetilde\Lambda}_\mathcal{F}$ eigenstates of 1D PXP-type PBC chains with extended Rydberg blockades have similar TI MPS forms, but the bond dimension increases with the blockade radius.
In higher dimensions, these eigenstates can be efficiently represented as projected entangled pair states (PEPS) \suppcite{Cirac_2021}.

\subsection{Exact entanglement entropy of entanglement-minimizing bipartitions}
\begin{figure}
  \subfloat{\label{fig:cut0}}
  \subfloat{\label{fig:cut1}}
  \subfloat{\label{fig:cut1equiv}}
  \subfloat{\label{fig:eedang}}
  \subfloat{\label{fig:eering}}\par\nointerlineskip
  \includegraphics[width=\textwidth]{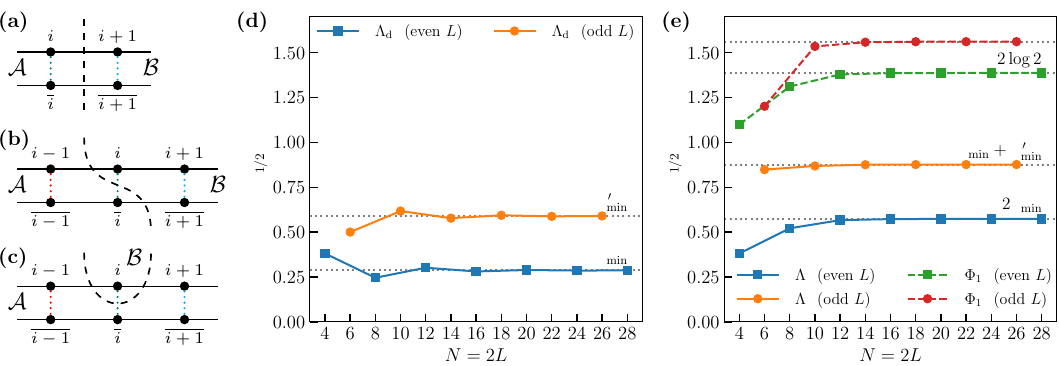}
  \caption{
    Entanglement-minimizing cuts and convergence of the half-system entanglement entropy to its exact asymptotic values for even and odd $L$.
    (a)~Regular entanglement-minimizing cut in the MPS chain.
    (b)~ A cut in the MPS chain splitting a pair of perfectly correlated spins.
    This type of a cut is unavoidable in entanglement-minimizing bipartitions into identical half-systems when $L$ is odd.
    (c)~A cut with one of the subsystems ($\mathcal{B}$) consisting of a single site. For $\ket{\Lambda}$ and $\ket{\Lambda_\text{d}}$ this cut has identical properties to the cut in (b).
    (d) Half-system entanglement entropy of the state $\ket{\Lambda_\text{d}}$.
    (e)~Half-system entanglement entropies of the currently known exact scars states of the PXP chain with PBCs.
    For the state $\ket{\Phi_1}$ from Ref.~\suppcite{Lin_2019} its entanglement-minimizing (i.e., standard) bipartition was used.
    For both even and odd $L$, our new state $\ket{\Lambda}$ can be bipartitioned into identical subsystems sharing even less entanglement than that for any bipartition of $\ket{\Phi_1}$.
  }
  \label{fig:cuts}
\end{figure}

Many otherwise difficult calculations are quite straightforward with MPS.
Let us calculate the exact eigenvalues of the reduced density matrices corresponding to the entanglement-minimizing bipartitions of $\ket{\Lambda}$ and $\ket{\Lambda_\text{d}}$ in the thermodynamic limit (i.e., when both subsystems are very large) and obtain the corresponding entropies of entanglement.
For $\ket{\Lambda}$, an example of such a bipartition cutting across four bonds is given in Fig.~(\ref{fig:bipart}); in general, given a PBC chain of size $2L$, four physical bonds --- namely, $(\dual L, 1)$, $(L, \dual 1)$, $(k, k+1)$ and $(\dual k, \dual{k+1})$, where $1 \ll k \ll L$ ---- are cut. For $\ket{\Lambda_\text{d}}$ of the same size, entanglement-minimizing bipartition cuts through two physical bonds: $(k,k+1)$ and $(\dual k, \dual{k+1})$.
Note that in terms of the MPS basis we used for $\ket{\Lambda}$ and $\ket{\Lambda_\text{d}}$ two physical cuts $(i, i+1)$ and $(\dual i, \dual{i+1})$ [shown in Fig.~\ref{fig:cut0}] amount to a single cut in the MPS chain.
Thus, the entanglement-minimizing bipartitions for $\ket{\Lambda_\text{d}}$ and $\ket{\Lambda}$ amount to, respectively, one and two cuts splitting the corresponding MPS chains.
In the thermodynamics limit, the two cuts in the PBC MPS chain like the ones described above are infinitely far from each other, which means they are entropically equivalent to two cuts of the same kind, each in the middle of an OBC chain with the same bulk MPS.
This means the asymptotic entanglement entropy for the case of PBCs is simply the sum of the independent contributions from each of the cuts assumed in the middle of an infinitely long OBC chain.

Using standard MPS techniques like those based on Gram matrices employed in Refs.~\suppcite{Moudgalya_2018entanglement} and \suppcite{Lin_2019}, or the infinite Density Matrix Renormalization Group (iDMRG) approaches \suppcite{mcculloch2008infinitesizedensitymatrix}, one can readily obtain the eigenvalues of the RDM corresponding to either of the subsystems produced by a cut in the middle of an OBC chain.
These eigenvalues for $\ket{\Lambda_\text{d}}$ are:
\begin{equation}
  \{\lambda_1, \lambda_2\} = \left\{ \frac{1}{2} + \gamma, \frac{1}{2} -\gamma \right\},
\end{equation}
where
\begin{equation}
  \gamma = \sqrt{\frac{1}{2\sqrt{5}}-\frac{1}{20}}\approx 0.4167.
\end{equation}
Hence, the asymptotic entanglement entropy for the entanglement-minimizing bipartition of $\ket{\Lambda_\text{d}}$ is
\begin{equation}
  \mathcal{S}_\text{min} = -\sum_{\mathclap{p \in \{\lambda_1, \lambda_2\}}}p\log p \approx 0.2868.
\end{equation}
On the other hand, it can be argued that the asymptotic eigenvalues corresponding to the RDMs resulting from an entanglement-minimizing bipartition of $\ket{\Lambda}$ are $\{\lambda_1, \lambda_2\}^{\times 2}$, where $\times$ denotes direct product (since there are two identical cuts in the MPS chain); hence the corresponding entanglement entropy is $2\mathcal{S}_\text{min}\approx 0.5736$.

Of particular interest are half-system entanglement-minimizing bipartitions splitting the system into two identical halves.
Note that the above analysis works for such bipartition only when $L$ is even (since we need $k=L/2$).
When $L$ is odd, it is impossible to split the system into two identical halves without separating a pair of perfectly correlated spins such that spin $i$ is placed into one subsystem and $\dual i$ is placed into another.
A cut of this type in our MPS chain is shown in Fig.~\ref{fig:cut1}.
Its properties are easiest to obtain indirectly by recognizing that physical bonds $(i, i+1)$ and $(\dual i, \dual{i+1})$ are identical for $\ket{\Lambda}$ and $\ket{\Lambda_\text{d}}$, which means the cuts in Figs.~\ref{fig:cut1} and \ref{fig:cut1equiv}, assuming an OBC MPS chain, result in exactly the same RDMs for the corresponding subsystems $\mathcal{A}$ and $\mathcal{B}$ (this visual argument can be verified by explicitly writing Schmidt decompositions for both situations, assuming the cuts are made in the middle of an OBC chain of some fixed size, and showing that the corresponding two non-zero Schmidt values must match).
Hence, we can immediately read-off the following asymptotic eigenvalues of the RDMs for the cut in Fig.~\ref{fig:cut1} from Eq.~(\ref{eq:rho1inf}):
\begin{equation}
  \{\lambda_1', \lambda_2'\} = \left\{\frac{\varphi}{\sqrt{5}}, \frac{1}{\sqrt{5}\varphi}\right\}.
\end{equation}
This gives
\begin{equation}
  \mathcal{S}_\text{min}' = -\sum_{\mathclap{p \in \{\lambda_1', \lambda_2'\}}}p\log p \approx 0.5895.
\end{equation}
In Figs.~\ref{fig:eedang} and \ref{fig:eering} we confirm numerically that asymptotic entropies of entanglement-minimizing bipartitions splitting  $\ket{\Lambda_\text{d}}$ and $\ket{\Lambda}$, respectively, into two identical halves are expressed exactly in terms of $\mathcal{S}_{\min}$ and $\mathcal{S}_{\min}'$: $\mathcal{S}_{\min}$ and $\mathcal{S}_{\min}'$ for $L$ even and odd in the $\ket{\Lambda_\text{d}}$ case, and $2\mathcal{S}_{\min}$ and $\mathcal{S}_{\min} + \mathcal{S}_{\min}'$ for $L$ even and odd in the $\ket{\Lambda}$ case.

\section{Graph modifying operator}
\label{supp:wop}
\begin{figure}
  \includegraphics[width=\textwidth]{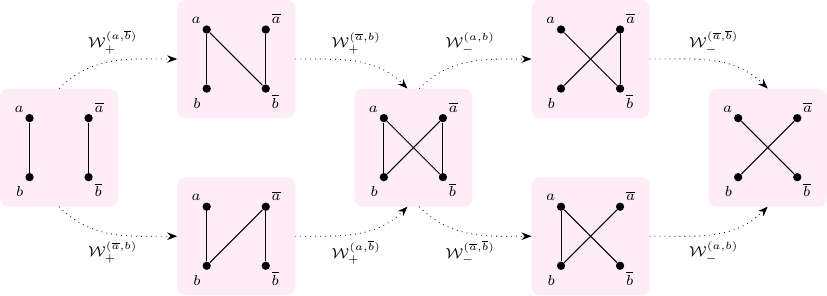}
  \caption{
    Possible patterns of couplings that can be obtained by actions of the graph modifying operator in Eq.~(\suppref{eq:gc}).
    Vertex pairs $(a, b)$ and $(\dual a, \dual b)$ are assumed to belong to originally isomorphic graphs $G_\mathcal{A}$ and $G_\dualmathcal{A}$, respectively.
    The actions shown together with their inverses (obtained by flipping the direction of arrows and exchanging $+$ and $-$ subscripts) exhaust all non-trivial possibilities.
  }
  \label{fig:wop}
\end{figure}

Here we provide visualizations of the actions of the graph modifying operator used to produce geometric generalizations in the main text.
In Fig.~\ref{fig:wop} we focus on two pairs of vertices representing interacting spins in subsystems $\mathcal{A}$ and $\dualmathcal{A}$ and show how the graph modifying operator defined in Eqs.~(\suppref{eq:gc}) and (\suppref{eq:gmo}) can be used to generate couplings between the subsystems that preserve the eigenstate $\ket*{\widetilde\Lambda}_\mathcal{F}$ hosted by the Hamiltonian $\widetilde H_\text{PXP}(G_\mathcal{A} \cup G_\dualmathcal{A})$ describing two identical uncoupled subsystems.

\section{No \texorpdfstring{$\ket*{\widetilde\Lambda}_\mathcal{F}$}{Lambda-type} state on OBC chains}
\label{supp:noobc}
Suppose $G_\mathcal{C}$ is a chain of length $2L$ with OBCs.
Note that any sequence of operators in Eq.~(\suppref{eq:gmo}) cannot decrease the initial degree of a vertex in $G_\mathcal{A} \cup G_\dualmathcal{A}$.
Hence one vertex in $G_\mathcal{A}$ must have a degree of 1, whereas all other vertices must have a degree of at most 2. Since obtaining the connected $G_\mathcal{C}$ is impossible unless $G_\mathcal{A}$ is connected, $G_\mathcal{A}$ must be an OBC chain consisting of $L$ sites; let these sites have labels $1,2,\dots,L$.
The only way to produce $G_\mathcal{C}$, then, is to combine $G_\mathcal{A}$ and $G_\dualmathcal{A}$ with a single edge $(a, \dual b) \in \{(1, \dual 1)$, $(1, \dual L)$, $(L, \dual 1)$, $(L, \dual L)\}$; i.e., by acting with $\mathcal{W}_+^{(a, \dual b)}$.
However, such $\mathcal{W}_+^{(a, \dual b)}$ could act only if edge $(a, b) \in \{(1, 1), (1, L), (L, 1), (L, L)\} $ is in $G_\mathcal{A}$, which (unless $L = 2$) is not possible.

\section{Uniqueness of \texorpdfstring{$\ket*{\widetilde\Lambda}_\mathcal{F}$}{state Lambda} on \texorpdfstring{$\mathcal{R}$}{R} from global pairing pattern}
\label{supp:uniqueallzz}
In this section we show that no two eigenstates on the Rydberg blockaded subspace $\mathcal{R}$ of $\widetilde H_\mathrm{PXP}(G)$ --- with $G=(V, E)$ being an arbitrary graph with $N=2L$ vertices --- can share the same global pairing pattern of perfectly correlated spins.
More precisely, given a pairing of the vertices of $G$ into $L$ non-overlapping pairs, which we will denote by $(a, b)_G$, there can exist at most one eigenstate of $\widetilde H_\text{PXP}(G)$ on $\mathcal{R}$ characterized by it.
Let $G_\mathcal{C} \cong G$ be a graph isomorphic to $G$ with vertices labeled using a bijection $(i, \dual i)_{G_\mathcal{C}} \leftrightarrow (a, b)_{G}$ relating labels of $G_\mathcal{C}$ and $G$.
Then, clearly, $\widetilde H_\text{PXP}(G)$ hosts an eigenstate with correlated pairs $(a, b)_G$ iif  $\widetilde H_\text{PXP}(G_\mathcal{C})$ hosts an eigenstate with correlated pairs $(i, \dual i)_{G_\mathcal{C}}$.
Hence, w.l.o.g.\ we will assume a graph $G_\mathcal{C}$ and a pairing pattern $(i, \dual i)_{G_\mathcal{C}}$ containing $L$ non-overlapping pairs that cover all vertices of $G_\mathcal{C}$.

\begin{lemma}
  \label{lemma:localpxp}
  If $\ket{\psi} \in \mathcal{R}$ is a simultaneous eigenstate of $\widetilde H_\mathrm{PXP}(G_\mathcal{C})$ and $Z_iZ_{\dual i}$ for a given $i$, then $\ket{\psi}$ must be annihilated by $\mathcal{P}_\mathcal{R}(X_i + X_{\dual i})$, where $\mathcal{P}_\mathcal{R}$ is a projector onto $\mathcal{R}$, or, equivalently, $\ket{\psi}$ must be annihilated by the the sum of two terms of $\widetilde H_\mathrm{PXP}(G_\mathcal{C})$ that have an $X$ operator acting on spins at sites $i$ or $\dual i$.
\begin{proof}
  If $\ket{\psi}$ is a simultaneous eigenstate of $\widetilde H_\mathrm{PXP}(G_\mathcal{C})$ and $Z_iZ_{\dual i}$, it must be annihilated by
\begin{equation}
  [\widetilde H_\mathrm{PXP}(G_\mathcal{C}), Z_iZ_{\dual i}]
  = \mathcal{P}_\mathcal{R}([X_i, Z_i]Z_{\dual i} + Z_i[X_{\dual i}, Z_{\dual i}]) \propto
  \mathcal{P}_\mathcal{R}(Y_iZ_{\dual i} + Z_iY_{\dual i}),
\end{equation}
and hence by
\begin{equation}
  Z_iZ_{\dual i}[\widetilde H_\mathrm{PXP}(G_\mathcal{C}), Z_iZ_{\dual i}] \propto \mathcal{P}_\mathcal{R}(X_i +
  X_{\dual i}).
\end{equation}
In the above manipulations we used the fact that $[\mathcal{P}_\mathcal{R}, Z_iZ_{\dual i}] = 0$.
\end{proof}
\end{lemma}

\begin{theorem}
  \label{thrm:uniqueness}
  If $\ket{\psi} \in \mathcal{R}$ is an eigenstate of $\widetilde H_\mathrm{PXP}(G_\mathcal{C})$ and
  \begin{equation}
    \label{eq:pairing}
    Z_iZ_{\dual i}\ket{\psi} = \ket{\psi} \text{~for~}\ \forall i \in [1..L], 
  \end{equation}
  it must have zero energy and be unique.
  \begin{proof}
    Eq.~(\ref{eq:pairing}) restricts $\ket{\psi}$ to the form
  \begin{equation}
    \label{eq:psiform}
    \ket{\psi} = \sum_{\ket{f} \in \mathcal{K}}c_f\ket{f}\otimes\ket{f},
  \end{equation}
  where $c_f$ are some coefficients and $\mathcal{K} \subset \mathbb{C}^{2^L}$ is a subspace spanned by computational basis states $\ket{f}$ such that $\ket{f}\otimes\ket{f} \in \mathcal{R}$.
  Thus,
  $\mathcal{C}_\text{ph}\ket{\psi} = \ket{\psi}$ --- i.e., $\ket{\psi}$ has definite parity.
  This is only possible if $\ket{\psi}$ has zero energy.

  Per Lemma~\ref{lemma:localpxp}, $\ket{\psi}$ must be annihilated by
  $\mathcal{P}_\mathcal{R}(X_i + X_{\dual i})$ for $\forall i \in [1..L]$, which gives
  \begin{equation}
    \label{eq:annihilators}
    \mathcal{P}_\mathcal{R}(X_i+X_{\dual i})\ket{\psi}
    = \mathcal{P}_\mathcal{R}\sum_{\ket{f} \in \mathcal{K}}
    c_f 
    \big[
    \left(X_i\ket{f}\right)\otimes\ket{f} + \ket{f}\otimes\left(X_{\dual i}\ket{f}\right) 
    \big]
    = 0.
  \end{equation}
  Taking an inner product of both sides Eq.~(\ref{eq:annihilators}) with an arbitrary product state  $\bra{s}\otimes\bra{t} \in \mathcal{R}$ we get
  \begin{equation}
    \label{eq:innerprod}
    \sum_{\ket{f}\in\mathcal{K}}\left(
      c_f\mel{s}{X_i}{f}\braket{t}{f} + c_f\braket{s}{f}\mel{t}{X_i}{f}
    \right) = 0,
  \end{equation}
  where we used $X_i$ instead of $X_{\dual i}$ since the labels are arbitrary in expressions involving a half-system. Equation~(\ref{eq:innerprod}) can be written as
  \begin{equation}
    \label{eq:ctcs}
    c_t\mel{s}{X_i}{t} + c_s\mel{t}{X_i}{s} = (c_s + c_t)\mel{s}{X_i}{t} = 0.
  \end{equation}
  A somewhat subtle point is that in going from Eq.~(\ref{eq:innerprod}) to Eq.~(\ref{eq:ctcs}) we implicitly restricted $\ket{s}$ and $\ket{t}$ to subspace $\mathcal{K}$.
  This is, of course, not problematic since given the form of Eq.~(\ref{eq:psiform}) $\mathcal{K}$ is exactly the subspace of interest.
  
  Suppose $|t|$ is the Hamming weight (number of ones in bitstring $t$) of some $\ket{t}\in\mathcal{K}$.
  Consider the set $\Omega^-_t = \left\{\ket{q}: \ket{q} = X_i\ket{t} \ \text{and}\ |q| = |t| - 1\right\}$. 
  Clearly $\Omega^-_t = \varnothing$ iif $\ket{t} = \ket{00\ldots 0}$; otherwise, $\Omega^-_t$ is a nonempty set and $\Omega^-_t \subset \mathcal{K}$.
  This means for $\forall\ket{t} \in \mathcal{K}$, $|t|>0$ $\exists \ket{q_1} \in \Omega^-_t$, which then implies $c_{q_1} + c_t = 0$. 
  But then if $|q_1| > 0$, $\exists \ket{q_2} \in \Omega^-_{q_1}$, implying $c_{q_2} + c_{q_1} = 0$, and so on.
  This recurrence terminates at $|q_{|t|-1}|= 1$ yielding $c_{q_{|t|}} + c_{q_{|t|-1}} = 0$, where $q_{|t|} = 00\ldots 0$. 
  Using these relations we obtain
  \begin{equation}
    \label{eq:cf}
    c_t = (-1)^{|t|}\cdot c_{00\dots 0}.
  \end{equation}
  Since $\ket{t} \in \mathcal{K}$ was arbitrary, we conclude that every amplitude in Eq.~(\ref{eq:psiform}) is related to $c_{00\dots 0}$.
  Thus, after plugging $c_t$ given in Eq.~(\ref{eq:cf}) into Eq.~(\ref{eq:psiform}) and normalizing we get a unique state formally identical to $\ket*{\widetilde\Lambda}$ given in Eq.~(\suppref{eq:lambdagen}) of the main text.
  \end{proof}
\end{theorem}

We note that the above argument for the uniqueness of the $\ket*{\widetilde\Lambda}_\mathcal{F}$ state for a given global pairing pattern holds for any PXP-type model on any graph $G_C$, but it does not guarantee that the state exists as an eigenstate (hence we had to assume it does).
By examining the proof we can also formulate a simple sufficient condition for the eigenstate to exist in the Rydberg-blockaded subspace $\mathcal{R}$ of $\widetilde H_\text{PXP}(G_\mathcal{C})$ referencing only the $G_C$ and the global pairing $(i, \dual i)$, which, naturally, defines two subgraphs $\widetilde G_\mathcal{A}$ and $\widetilde G_\dualmathcal{A}$ with the same number of vertices but not necessarily the same structure.
We also assume that the graph $G_C$ does not contain edges connecting $i$ and $\dual i$ for every $i$ (otherwise the corresponding spins would be fixed as $\ket{00}_{i\dual i}$ in the $\ket*{\widetilde\Lambda}_\mathcal{F}$ state, which is not interesting).
Consider sets $\Omega_i = \{j, (i,j) \in E\}$ and $\Omega_{\dual i} = \{j, (\dual i,j) \in E\}$ for $\forall i$.
Given an arbitrary $G_\mathcal{C}$, these sets can contain vertices belonging to both subgraphs $\widetilde G_\mathcal{A}$ and $\widetilde G_\dualmathcal{A}$. Let us introduce function $u(v)$ that acts as identity on the vertices $\widetilde G_\mathcal{A}$, but maps vertices of $\widetilde G_\dualmathcal{A}$ to the corresponding vertices of $\widetilde G_\mathcal{A}$ (e.g., $u(\dual 1)$ = $1$, but $u(1) = 1$).
Now, a sufficient condition for the existence of a $\ket*{\widetilde\Lambda}_\mathcal{F} \in \mathcal{R}$ state given a global pairing pattern is the following:
\begin{equation}
\{u(j), j \in \Omega_i\} = \{u(j), j \in \Omega_{\dual i}\} \equiv \widetilde\Omega_i.
\label{eq:graphcondition}
\end{equation}

In the wavefunction amplitude language as in the above uniqueness proof, this condition implies that, for any $i$ and any $\ket{g}_{\text{rest}}$, we have two non-trivial possibilities:
either both $\ket{00}_{i\dual i}\ket{g}_{\text{rest}} \ket{g}_{\dual{\text{rest}}}$ and $\ket{11}_{i\dual i}\ket{g}_{\text{rest}} \ket{g}_{\dual{\text{rest}}}$ are in $\mathcal{R}$ and hence contribute to $\ket*{\widetilde\Lambda}$ with opposite amplitudes, in which case their total contribution is annihilated by $\mathcal{P}_{\mathcal{R}} (X_i + X_{\dual{i}})$;
or $\ket{00}_{i\dual i}\ket{g}_{\text{rest}} \ket{g}_{\dual{\text{rest}}}$ is in $\mathcal{R}$ but $\ket{11}_{i\dual i}\ket{g}_{\text{rest}} \ket{g}_{\dual{\text{rest}}}$ is not due to blockades at $i$ and $\dual i$ provided by the rest of the spins, in which case only the former contributes to $\ket*{\widetilde\Lambda}$ and this contribution is annihilated by both $\mathcal{P}_{\mathcal{R}} X_i$ and $\mathcal{P}_{\mathcal{R}} X_{\dual{i}}$ due to the same blockades by the rest as guaranteed by the above condition.
Hence $\mathcal{P}_{\mathcal{R}} (X_i + X_{\dual{i}}) \ket*{\widetilde\Lambda} = 0$.

In the graph language, the condition of Eq.~(\ref{eq:graphcondition}) effectively says that $G_\mathcal{C}$ can be constructed from two isomorphic graphs $G_\mathcal{A}$ and $G_\dualmathcal{A}$ (sharing all the vertices, but not necessarily edges, with, respectively,  $\widetilde G_\mathcal{A}$ and $\widetilde G_\dualmathcal{A}$) via the construction of Eq.~(\suppref{eq:gc}) in the main text. Given the condition is satisfied, we can define graph $G_\mathcal{A}$ explicitly as follows: $G_\mathcal{A} = (V_\mathcal{A}, \{(i,j), i \in V_\mathcal{A}\ \text{and}\ j \in \widetilde\Omega_i\})$, where $V_\mathcal{A}$ are the vertices of $\widetilde G_\mathcal{A}$. Then the said $\ket*{\widetilde\Lambda}_\mathcal{F}$ state will purify the maximally mixed ensemble on $\mathcal{R}$ of $\widetilde H_\text{PXP}(G_\mathcal{A})$ on the system with the Hamiltonian $\widetilde H_\text{PXP}(G_\mathcal{C})$.

\section{Uniqueness of the eigenstate of \texorpdfstring{$Z_{1}Z_{\dual 1}$}{Z1Z1bar} on \texorpdfstring{$\mathcal{R}$}{R} on the dangler configuration}
\label{supp:dangler}
Our goal in this section is to show that there is only one eigenstate of the PXP model on the dangler geometry that is also a $+1$ eigenstate of $Z_1 Z_{\dual 1}$, namely the $\ket{\Lambda_\text{d}}$ state from the main text.
Note that this is a much stronger statement than in the preceding section since we are requiring that only one pair of sites, $1$ and $\dual 1$, is correlated.

Suppose $G_\mathcal{C}$ is the dangler graph [as in Fig.~\suppref{fig:dangler} of
the main text] with $N=2L$ vertices, and $H_\text{d} = \widetilde H_\text{PXP}(G_\mathcal{C})$. The proof, established mainly via the following two lemmas, will lie in showing that any $+1$ eigenstate of $Z_1Z_{\dual 1}$ must be characterized by a global paring pattern.

\begin{lemma}
  \label{lemma:z1z1}
  If $\ket{\psi}$ is an eigenstate of $H_\mathrm{d}$ and $Z_1Z_{\dual 1}\ket{\psi} = \ket{\psi}$ then $\ket{\psi}$ must be a $+1$ eigenstate of $Z_2Z_{\dual 2}$.
  \begin{proof}
    Per Lemma~\ref{lemma:localpxp}, $\ket{\psi}$ must be annihilated by
    \begin{equation}
      H_1 = P_2X_1P_{\dual 2} + X_{\dual 1}P_{\dual 2}.
    \end{equation}
    Note that at this point we cannot assume anything about the relationship between spins at sites $2$ and $\dual 2$.
    
    Let us write $\ket{\psi}$ as
    \begin{equation}
      \label{eq:phiz1z1}
      \ket{\psi} = \sum_{s_2s_1s_{\dual 2} \in \mathcal{F}^{\text{OBC}}_3}
      \ket{s_1s_1}_{1,\dual 1}\ket{s_2s_{\dual 2}}_{2,\dual 2}\ket*{\tilde\phi_{s_2s_1s_{\dual 2}}}_\text{rest},
    \end{equation}
    where $\ket*{\tilde\phi_{s_2s_1s_{\dual 2}}}_\text{rest}$ are unnormalized wavefunctions defined on the subsystem consisting of all spins except those with labels $1,\dual 1,2,\dual 2$.
    For conciseness, in what follows we will drop the subscripts indicating the subsystems on the kets in the above tensor product, always understanding them appearing in the same order as in the above.
    From
    \begin{equation}
      H_1\ket{\psi} = \ket{10}\ket{00}\ket*{\tilde\phi_{000}}
      + \ket{01}\ket{00}\ket*{\tilde\phi_{000}}
      +\ket{01}\ket{10}\ket*{\tilde\phi_{100}}
      +\ket{10}\ket{00}\ket*{\tilde\phi_{010}}
      +\ket{01}\ket{00}\ket*{\tilde\phi_{010}}= 0,
    \end{equation}
    we deduce that
    \begin{subequations}
      \begin{align}
        \label{eq:useless}
        & \ket*{\tilde\phi_{000}} = -\ket*{\tilde\phi_{010}}, \\
        \label{eq:phi100}
        &\ket*{\tilde\phi_{100}} = 0.
      \end{align}
    \end{subequations}
    While Eq.~(\ref{eq:useless}) gives us little useful information at this point, Eq.~(\ref{eq:phi100}) says that components with spins $2$ and $\dual 2$ in states, respectively, $\ket{1}$ and $\ket{0}$ are forbidden.
    Thus the only term in Eq.~(\ref{eq:phiz1z1}) that is not a $+1$ eigenstate of $Z_2Z_{\dual 2}$ is the one with $s_2s_1s_{\dual 2} = 001$.
    To prove the Lemma we need to show that it must vanish as well.

    Let us write $\ket{\psi}$ explicitly utilizing everything we have learned about its form so far:
    \begin{equation}
      \label{eq:psiexplicit}
      \ket{\psi} = \ket{00}\left(\ket{00}\ket*{\tilde\phi_{000}} + \ket{01}\ket*{\tilde\phi_{001}}
        + \ket{11}\ket*{\tilde\phi_{101}} \right) - \ket{11}\ket{00}\ket*{\tilde\phi_{000}},
    \end{equation}
    where we used the result of Eq.~(\ref{eq:useless}) to write the last term.
    Consider the individual unnormalized terms in Eq.~(\ref{eq:psiexplicit}):
    \begin{subequations}
      \begin{align}
        \ket*{\tilde\psi_{00}} &= \ket{00}\left(\ket{00}\ket*{\tilde\phi_{000}} + \ket{01}\ket*{\tilde\phi_{001}} + \ket{11}\ket*{\tilde\phi_{101}}\right), \\
        \ket*{\tilde\psi_{11}} &= \ket{11}\ket{00}\ket*{\tilde\phi_{000}}.
      \end{align}
    \end{subequations}
    Thus,
    \begin{equation}
      \ket{\psi} = \ket*{\tilde\psi_{00}} - \ket*{\tilde\psi_{11}}.
    \end{equation}

    Now, $\ket{\psi}$ must be an eigenstate of the Hamiltonian
    \begin{equation}
      H' = H_\text{d} - H_1
    \end{equation}
    with the same eigenvalue $\lambda$ as that in $H_\text{d}\ket{\psi} = \lambda\ket{\psi}$.
    Since $H'$ doesn't modify spins $1$ and $\dual 1$, and therefore doesn't mix $\ket*{\tilde\psi_{00}}$ and $\ket*{\tilde\psi_{11}}$, we conclude that
    \begin{subequations}
      \begin{align}
        \label{eq:psi00}
        H'\ket*{\tilde\psi_{00}} &= \lambda\ket*{\tilde\psi_{00}}, \\
        \label{eq:psi11}
        H'\ket*{\tilde\psi_{11}} &= \lambda\ket*{\tilde\psi_{11}}.
      \end{align}
    \end{subequations}
    Let us rewrite Eq.~(\ref{eq:psi11}) explicitly as follows:
    \begin{equation}
      \label{eq:hbarpsi11}
      H'\ket*{\tilde\psi_{11}} = \ket{11}\left[
        \left(H' - H_2\right)\ket{00}\ket*{\tilde\phi_{000}}
      \right]
      = \lambda\ket{11}\ket{00}\ket*{\tilde\phi_{000}},
    \end{equation}
    where $H_2$ denotes terms of $H'$ with $X$ operators
    acting on sites $2$ and $\dual 2$; i.e.,
    \begin{equation}
      H_2 = P_3X_2P_1 + P_{1}P_{\dual 1}X_{\dual 2}P_{\dual 3}.
    \end{equation}
    From Eq.~(\ref{eq:hbarpsi11}) --- which is correct because
    $H_2\ket*{\tilde\psi_{11}} = 0$ and $H' - H_2$ has no
    support on sites $1$ and $\dual 1$ -- we deduce that
    \begin{equation}
      \label{eq:psi000h2}
      \left(H' - H_2\right)\ket{00}_{2,\dual 2}\ket*{\tilde\phi_{000}}
      = \lambda\ket{00}_{2,\dual 2}\ket*{\tilde\phi_{000}}.
    \end{equation}
    Note that the wavefunction in Eq.~(\ref{eq:psi000h2}) is defined
    on the subsystem that excludes sites $1$ and $\bar 1$.

    Let us now write the action of $H'$ on
    $\ket*{\tilde\psi_{00}}$ in a rather strange fashion for reasons
    to become apparent shortly:
    \begin{align}
      \label{eq:hbarpsi00}
      H'\ket*{\tilde\psi_{00}} &= H_2\ket{00}\ket{00}\ket*{\tilde\phi_{000}}\\
      \label{eq:hbarpsi00term1}
                                        &+ \ket{00}\left[
                                          \left(H' - H_2\right)\ket{00}\ket*{\tilde\phi_{000}} \right]\\
      \label{eq:hbarpsi00term2}
                                        &+ H_2\ket{00}\ket{01}\ket*{\tilde\phi_{001}}\\
                                        &+ \ket{00}\left[
                                          \left(H' - H_2\right)\ket{01}\ket*{\tilde\phi_{001}} \right]\\
                                        &+ H'\ket{00}\ket{11}\ket*{\tilde\phi_{101}}.
    \end{align}
    Acting with projector $\mathcal{P}_{0000} = \ketbra{0000}{0000}_{1\dual 12\dual 2}$ on
    both sides of Eq.~(\ref{eq:psi00}), we get
    \begin{equation}
      \label{eq:projhbarpsi00}
      \mathcal{P}_{0000} H'\ket*{\tilde\psi_{00}} = \lambda \ket{00} \ket{00} \ket*{\tilde\phi_{000}}.
    \end{equation}
    Now, given the form of Eq.~(\ref{eq:hbarpsi00}), it is clear that
    $\mathcal{P}_{0000}$ will annihilate all terms except those in
    lines labeled as Eqs.~(\ref{eq:hbarpsi00term1}) and
    (\ref{eq:hbarpsi00term2}). Thus
    \begin{equation}
      \label{eq:projhbarpsi00expl}
      \mathcal{P}_{0000}  H'\ket*{\tilde\psi_{00}} = \ket{00}\left[
        \left(H' - H_2\right)\ket{00}\ket*{\tilde\phi_{000}} \right] + \ket{00}\ket{00}\ket*{\tilde\phi_{001}}.
    \end{equation}
    where in the last line we used $P_{\dual 3} \ket*{\tilde\phi_{001}} = \ket*{\tilde\phi_{001}}$ by Rydberg blockade on $\dual 3$ from excited state at $\dual 2$.
    Combining Eqs.~(\ref{eq:psi000h2}), (\ref{eq:projhbarpsi00}), and (\ref{eq:projhbarpsi00expl}), we obtain
    \begin{equation}
      \lambda\ket{00}\ket{00}\ket*{\tilde\phi_{000}} + \ket{00}\ket{00}\ket*{\tilde\phi_{001}} = \lambda\ket{00}\ket{00}\ket*{\tilde\phi_{000}} \implies \ket*{\tilde\phi_{001}} = 0.
    \end{equation}
    In the above analysis we silently assumed that $L > 2$.
    In the special case when $L = 2$, $H' - H_2 = 0$, $\ket*{\tilde\phi_{s_2s_1s_{\dual 2}}}_\text{rest}$ are just numbers, and one arrives at exactly the same conclusion following similar steps.    
  \end{proof}
\end{lemma}

\begin{lemma}
  \label{lemma:zzplus}
  If $\ket{\psi}$ is an eigenstate of $H_\mathrm{PXP}$ and $Z_iZ_{\dual i}\ket{\psi}=\ket{\psi}$ for $\forall i \in [1,\dots,m]$, where $2 \leq m < L$, then $\ket{\psi}$ must also be a $+1$ eigenstate of $Z_{m+1}Z_{\dual{m+1}}$.
\begin{proof}
  First, consider the case of $m=2$. We want to show that $\ket{\psi}$
  must be an eigenstate of $Z_3Z_{\dual 3}$ ($L\geq 3$). Per
  Lemma~\ref{lemma:localpxp}, $\ket{\psi}$ must be annihilated by the
  following operators:
  \begin{subequations}
    \begin{align}
      \label{eq:localpxp1}
      &H_1 = P_2X_1P_{\dual 2} + X_{\dual 1}P_{\dual 2} \simeq P_2X_1 + X_{\dual 1}P_{\dual 2}, \\
      \label{eq:localpxp2}
      &H_2 = P_3X_2P_1 + P_{1}P_{\dual 1}X_{\dual 2}P_{\dual 3} \simeq P_3X_2P_1 + P_{\dual 1}X_{\dual 2}P_{\dual 3},
    \end{align}
  \end{subequations}
  where in the expressions to the right of $\simeq$, we removed redundant projectors as far as acting on $\ket{\psi}$ is concerned.

  Let us express $\ket{\psi}$ as
  \begin{equation}
    \label{eq:psicomp}
    \ket{\psi} = \sum_{
      s_1s_2\in\mathcal{F}^{\text{OBC}}_2
    }
    \ket{s_1s_1}_{1,\dual 1}\ket{s_2s_2}_{2,\dual 2}\ket*{\tilde\phi_{s_1s_2}},
  \end{equation}
  where $\ket*{\tilde\phi_{s_1s_2}}$ are unnormalized wavefunctions defined on the subsystem consisting of all spins except those with labels $1, \dual 1, 2, \dual 2$. 
  Note that of the three terms in Eq.~(\ref{eq:psicomp}) the one with $s_1s_2=01$ is already a $+1$ eigenstate of $Z_3Z_{\dual 3}$; i.e.,
  \begin{equation}
    \label{eq:z3z3psi01}
    Z_3Z_{\dual 3}\ket*{\tilde\phi_{01}} = \ket*{\tilde\phi_{01}}
  \end{equation}
  because spins $3$ and $\dual 3$ are forced to be in the state
  $\ket{0}$ by the adjacent spins $2$ and $\dual 2$ in the state
  $\ket{1}$.
  
  With spins in the same order as in Eq.~(\ref{eq:psicomp}), the
  action of $H_2$ on $\ket{\psi}$ can be written as
  \begin{equation}
    H_2\ket{\psi} = \ket{00}\left[\ket{01}\left(\ket*{\tilde\phi_{01}} + P_{\dual 3}\ket*{\tilde\phi_{00}}\right)
      + \ket{10}\left(\ket*{\tilde\phi_{01}} + P_{3}\ket*{\tilde\phi_{00}}\right)\right],
  \end{equation}
  where we have used $P_3 \ket*{\tilde\phi_{01}} = P_{\dual 3} \ket*{\tilde\phi_{01}} = \ket*{\tilde\phi_{01}}$.
  For $H_2$ to annihilate
  $\ket{\psi}$ the following two conditions must be satisfied:
  \begin{subequations}
    \begin{align}
      & P_3\ket*{\tilde\phi_{00}} = -\ket*{\tilde\phi_{01}},\\
      & P_{\dual 3}\ket*{\tilde\phi_{00}} = -\ket*{\tilde\phi_{01}}.
    \end{align}
  \end{subequations}
  Clearly, this is only possible if $\ket*{\tilde\phi_{00}}$ does not have any components where spins $3$ and $\dual 3$ are in different states.
  Thus we conclude that
  \begin{equation}
    \label{eq:z3z3psi00}
    Z_3Z_{\dual 3}\ket*{\tilde\phi_{00}} = \ket*{\tilde\phi_{00}}.
  \end{equation}
  
  Finally, acting with $H_1$ on $\ket{\psi}$ we get
  \begin{equation}
    H_1\ket{\psi} = (\ket{10}\ket{00} +
    \ket{01}\ket{00})(\ket*{\tilde\phi_{00}} + \ket*{\tilde\phi_{10}}),
  \end{equation}
  which gives $\ket*{\tilde\phi_{10}} = -\ket*{\tilde\phi_{00}}$ and, therefore,
  \begin{equation}
    \label{eq:z3z3psi10}
    Z_3Z_{\dual 3}\ket*{\tilde\phi_{10}} = \ket*{\tilde\phi_{10}}.
  \end{equation}
  Combining Eqs.~(\ref{eq:psicomp}), (\ref{eq:z3z3psi01}),
  (\ref{eq:z3z3psi00}), and (\ref{eq:z3z3psi10}) we conclude that
  \begin{equation}
    Z_3Z_{\dual 3}\ket{\psi} = \ket{\psi}.
  \end{equation}
  
  Suppose the conditions of the Lemma are satisfied up to some $m > 2$.
  Then $\ket{\psi}$ must be annihilated by operators $H_1, H_2, \dots, H_m$, analogous to the ones given in Eqs.~(\ref{eq:localpxp1}) and (\ref{eq:localpxp2}).
  We can express $\ket{\psi}$ as
  \begin{equation}
    \label{eq:psicompgen}
    \ket{\psi} = \sum_{s_1, s_2, \dots, s_m \in \mathcal{F}^{\text{OBC}}_m}\ket{s_1\ldots s_{m-2}}_{1,\ldots,m-2} \ket{s_1\ldots s_{m-2}}_{\dual 1,\ldots,\dual{m-2}}
    \ket{s_{m-1}s_{m-1}}_{m-1,\dual{m-1}} \ket{s_ms_m}_{m,\dual m}\ket*{\tilde\phi_{s_{m-1}s_{m}}^{s_1\ldots s_{m-2}}},
  \end{equation}
  where $\ket*{\tilde\phi_{s_{m-1}s_m}^{s_1s_2\dots s_{m-2}}}$ are unnormalized wavefunctions defined on the subsystem consisting of all spins except those with labels $1, \dual 1, 2, \dual 2,\dots, m, \dual m$. 
  Now, via arguments identical to the ones we used in the $m=2$ case applied to individual wavefunction parts with any fixed $s_1s_2\dots s_{m-2}$ one can show that
  \begin{equation}
    \label{eq:zmplus}
    Z_{m+1}Z_{\dual{m+1}}\ket*{\tilde\phi_{s_{m-1}s_m}^{s_1s_2\dots s_{m-2}}}
    = \ket*{\tilde\phi_{s_{m-1}s_m}^{s_1s_2\dots s_{m-2}}}.
  \end{equation}
  Specifically, if $s_{m-2} = 0$, the argument is entirely identical with $H_m$ replacing $H_2$ and $H_{m-1}$ replacing $H_1$; whereas if $s_{m-2} = 1$, then $s_{m-1}s_m$ can only take values $00$ and $01$, which means Eq.~(\ref{eq:zmplus}) follows from only requiring that $H_m$ annihilate $\ket{\psi}$.
\end{proof}
\end{lemma}

We are now ready to prove the main result:
\begin{theorem}
  \label{thrm:danglerz1z1}
  The simultaneous eigenspace of
  $\widetilde H_\text{d}$ and $Z_1Z_{\dual 1}$ on
  $\mathcal{R}$ is one-dimensional.
  \begin{proof}
    Per Lemma \ref{lemma:z1z1}, any state $\ket{\psi}$ in the
    simultaneous eigenspace of
    $\widetilde H_\text{d}$ and $Z_1Z_{\dual 1}$ must
    be an eigenstate of $Z_2Z_{\dual 2}$. Then, by induction on the
    result of Lemma~\ref{lemma:zzplus}, $\ket{\psi}$ must also be an
    eigenstate of $Z_iZ_{\dual i}$ for any $i \in [1,\dots,L]$. Hence by
    Theorem \ref{thrm:uniqueness}, $\ket{\psi}$ must have zero energy
    and be unique. Since a state $\ket{\Lambda_\text{d}}$ satisfying the conditions of the
    Theorem has been constructed explicitly in the main text, the
    unique $\ket{\psi}$ discussed herein exists.
  \end{proof}
\end{theorem}

\begin{figure}
  \includegraphics[width=0.5\textwidth]{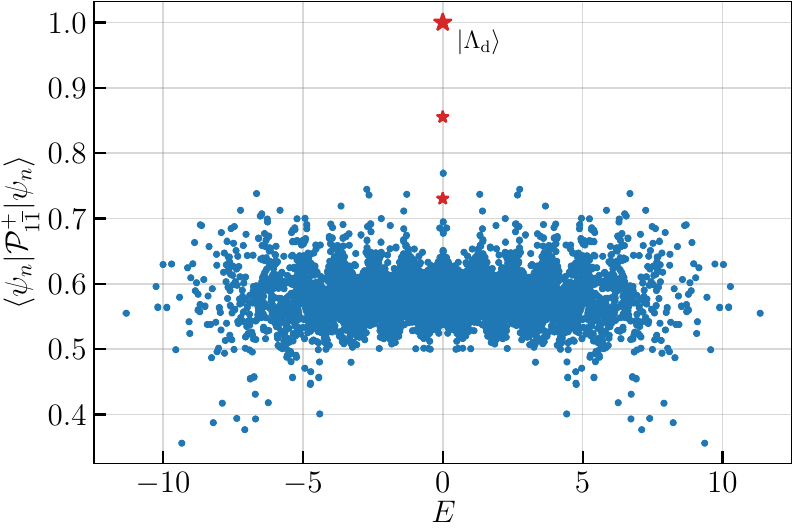}
  \caption{
    Squared projections of eigenstates of an $N = 18$ dangler system onto the $+1$ eigenspace of $Z_1Z_{\dual 1}$.
    Suppose normalized eigenstates of the Hamiltonian $H_\text{d}$ on $\mathcal{R}$ are written as $\ket{\psi_n} = \alpha_n \ket{\psi_n^+} + \beta_n \ket{\psi_n^-}$, where $\ket{\psi_n^{\pm}}$ are normalized projections onto the $\pm 1$ eigenspaces of $Z_1 Z_{\dual 1}$; the $+1$ projector is $\mathcal{P}^+_{1\dual 1} = \ketbra{00}{00}_{1\dual 1} + \ketbra{11}{11}_{1\dual 1}$.
    As expected per Theorem~\ref{thrm:danglerz1z1}, $\alpha_n = 1$ iif $\ket{\psi_n} = \ket{\Lambda_\text{d}}$; otherwise, $|\alpha_n| < 1$.    
    The smaller star markers indicate the other eigenstates in the three-fold degenerate nullspace ordered by the magnitude of the $\ket{\psi_n^+}$  component.
  }
  \label{fig:danglerzz}
\end{figure}

A numerical demonstration of Theorem~\ref{thrm:danglerz1z1} is given in Fig.~\ref{fig:danglerzz}.

\section{Uniqueness of the eigenstate of \texorpdfstring{$Z_{1}Z_{\dual 1}$}{Z1Z1bar} on two identical decoupled OBC chains}
In this section we establish the same result as that obtained in the previous section for the dangler, but for a simpler system consisting of two identical decoupled OBC chains.

Suppose $G_\mathcal{A}$ and $G_\dualmathcal{A}$ are isomorphic OBC chains of size $L$ (i.e., $G_\mathcal{A} \cup G_\dualmathcal{A}$ is the decoupled system that generates the dangler). Before we proceed, let us remark that Lemma~\ref{lemma:zzplus} applies without modification to the system considered here (one simply needs to assume the simplification made in Eq.~(\ref{eq:localpxp1}) from the start); hence, we will invoke Lemma~\ref{lemma:zzplus} in what follows.

\begin{theorem}
  \label{thrm:discobc}
  The simultaneous eigenspace of $\widetilde H_\text{PXP}(G_\mathcal{A} \cup G_\dualmathcal{A})$ and $Z_1Z_{\dual 1}$ on $\mathcal{R}$ is one-dimensional.
  \begin{proof}
    Per Lemma~\ref{lemma:localpxp}, $\ket{\psi}$ must be annihilated by
    \begin{equation}
      H_1 = P_2X_1 + X_{\dual 1}P_{\dual 2}.
    \end{equation}
    Consider $\ket{\psi}$ expressed in the form given by Eq.~(\ref{eq:phiz1z1}).
    From
    \begin{equation*}
      H_1 \ket{\psi} = \ket{10}\ket{00}\ket*{\tilde\phi_{000}}
      + \ket{10}\ket{01}\ket*{\tilde\phi_{001}}
      + \ket{01}\ket{00}\ket*{\tilde\phi_{010}}
      + \ket{01}\ket{00}\ket*{\tilde\phi_{000}}
      + \ket{01}\ket{10}\ket*{\tilde\phi_{100}}
      + \ket{10}\ket{00}\ket*{\tilde\phi_{010}} = 0,
    \end{equation*}
    we deduce that
    \begin{equation}
      \ket*{\tilde\phi_{000}} = -\ket*{\tilde\phi_{010}}, \quad
      \ket*{\tilde\phi_{001}} = \ket*{\tilde\phi_{100}} = 0.
    \end{equation}
    Hence $\ket{\psi}$ is a $+1$ eigenstate of $Z_2 Z_{\dual 2}$, which implies, by induction on the result of Lemma~\ref{lemma:zzplus}, that $\ket{\psi}$ is characterized by a global pairing pattern such that $Z_iZ_{\dual i}\ket{\psi} = \ket{\psi}$ for $\forall i \in [1,\dots,L]$.
    The rest of the argument is entirely identical to that used in Theorem~\ref{thrm:danglerz1z1}.
  \end{proof}  
\end{theorem}
In particular, this result implies that the same state $\ket{\Lambda_\text{d}}$, whose preparation was described in the main text using projective $Z_1 Z_{\dual 1}$ measurements and dynamics under the dangler PXP model, can also be prepared on decoupled OBC chains through an identical protocol (this is illustrated in Figs.~\ref{fig:dhinfidelity} and \ref{fig:dhpostsel}).

\section{Composite evolution and projective measurement operator \texorpdfstring{$M_\tau$}{M(t)}}
\label{supp:mtau}

In this section we argue that if there exists a provably unique state $\ket*{\widetilde\Lambda} \cong \ket*{\widetilde\Lambda}_\mathcal{F}$ that is a simultaneous eigenstate of some PXP-type Hamiltonian $H$ and projective measurement operator $\mathcal{P}^+_{1\dual 1} = \ketbra{00}{00}_{1\dual 1} + \ketbra{11}{11}_{1\dual 1}$ (in this discussion, labels $1$ and $\dual 1$ correspond to an arbitrary pair of correlated spins) then the non-Hermitian operator $M_\tau = \mathcal{P}^+_{1\dual 1} e^{-iH\tau}$, for generic $\tau > 0$, has the property that
\begin{equation}
  \label{eq:mtau}
  \lim_{k\to\infty} M_\tau^k = \ketbra*{\widetilde\Lambda}{\widetilde\Lambda}.
\end{equation}
That is, postselection on the $+1$ outcomes of a sequence of $Z_1 Z_{\dual 1}$ measurements preceded by evolution under $H$ over period $\tau$ (we will refer to measurements resulting in $+1$ outcomes as ``successful'') converges to the projector onto the $\ket*{\widetilde\Lambda}_\mathcal{F}$ eigenstate.
The discussion easily generalizes to the case of multiple states $\ket*{\widetilde\Lambda_1}, \ket*{\widetilde\Lambda_2}, \dots, \ket*{\widetilde\Lambda_n}$ with the same properties.

Let us define the postselection probability $p_k$ (i.e., the probability of obtaining $k$ consecutive successful measurements) as
\begin{equation}
  \label{eq:postseldef}
  p_k = \mel{\psi_{k-1}}{M_\tau^\dagger M_\tau}{\psi_{k-1}} \, p_{k-1},
\end{equation}
where $\ket{\psi_k} = M_\tau^k\ket{\psi_0} / \lVert M_\tau^k\ket{\psi_0} \rVert$ is the normalized wavefunction after $k$ successful measurements when one starts from the initial state $\ket{\psi_0}$.
The first factor in Eq.~(\ref{eq:postseldef}) is the conditional probability of a successful $k^\text{th}$ measurement given that the previous $k-1$ measurements have been successful. As expected for repeated measurements, the recurrence in Eq.~(\ref{eq:postseldef}) gives $p_k = \mel{\psi_0}{\left[M_\tau^\dagger\right]^k M_\tau^k}{\psi_0}$, which means that per Eq.~(\ref{eq:mtau}) the overall postselection probability is expected to converge to the squared overlap of the initial state with $\ket*{\widetilde\Lambda}$; i.e., $\lim_{k\to\infty} p_k = |\braket*{\widetilde\Lambda}{\psi_0}|^2$.

We first describe an intuitive argument for Eq.~(\ref{eq:mtau}).

Due to its non-Hermiticity, operator $M_\tau$ is not particularly easy to work with.
We can, however, express $M_\tau^k$ as a product of $k$ Hermitian operators and a unitary using the following structure:
\begin{equation}
  M_\tau^k = e^{-kiH\tau} \times e^{kiH\tau}\mathcal{P}_{1\dual 1}^+e^{-kiH\tau} \times \cdots \times e^{2iH\tau}\mathcal{P}_{1\dual 1}^+e^{-2iH\tau} \times e^{iH\tau}\mathcal{P}_{1\dual 1}^+e^{-iH\tau}.
\end{equation}
Note that in the context of amplitude damping (of anything orthogonal to $\ket{\Lambda}$) by $M_\tau^k$ the leftmost unitary $e^{-kiH\tau}$ has no effect and thus can be ignored.
We can, therefore, view $M_\tau^k$ as a product of Hermitian Heisenberg operators $\mathcal{P}_{1\dual 1}^+(t) = e^{iHt}\mathcal{P}_{1\dual 1}^+e^{-iHt}$ written as
\begin{equation}
  \label{eq:mtaukheis}
  M_\tau^k \simeq \mathcal{P}_{1\dual 1}^+(k\tau)\cdots\mathcal{P}_{1\dual 1}^+(2\tau)\mathcal{P}_{1\dual 1}^+(\tau).
\end{equation}

Clearly, $\sigma_{\mathcal{P}_{1\dual 1}^+(t)} = \sigma_{\mathcal{P}_{1\dual 1}^+}$, where $\sigma_T$ denotes the spectrum of an operator $T$; i.e., the spectra of all the operators in Eq.~(\ref{eq:mtaukheis}) are identical and contain only 0's and 1's.
Let $V_t^\lambda$, where $\lambda \in \{0,1\}$, be the time-dependent eigenspace of $\mathcal{P}_{1\dual 1}^+(t)$ corresponding to eigenvalue $\lambda$, $V_t^\lambda = e^{iHt} V_0^\lambda$; therefore, $P_{1\dual 1}^+(t)$ is a projector $\mathcal{H} \to V_{t}^1$, where $\mathcal{H} = V_t^0 \oplus V_t^1$ is the full Hilbert space.
For any two times $t_1$ and $t_2$, and any $\lambda$,
\begin{equation}
  \label{eq:heisprojeigenspace}
  V_{t_1}^\lambda\cap V_{t_2}^\lambda \subseteq V_{t_1}^\lambda,\quad V_{t_1}^\lambda\cap V_{t_2}^\lambda \subseteq V_{t_2}^\lambda
\end{equation}
Given the overall chaotic nature of the PXP Hamiltonian and only a single common eigenstate with $P_{1\dual 1}^+$, it is reasonable to assume that for generic $t_1 \neq t_2$, $V_{t_1}^\lambda\cap V_{t_2}^\lambda$ will be a proper subset of both $V_{t_1}^\lambda$ and $V_{t_2}^\lambda$.
In other words, subspaces $V_{t_1}^\lambda$ and $V_{t_2}^\lambda$ will not be exactly equal; and if the interval between $t_1$ and $t_2$ is sufficiently large, these subspaces can be considered to occupy random regions of the full Hilbert space.
Although each operator in Eq.~(\ref{eq:heisprojeigenspace}) has exactly $\dim V_0^1$ fixed points with eigenvalue 1, there is only one fixed point which is guaranteed to be shared among all these operators, namely $\ket{\Lambda}$.
This, together with the assumption that equality is almost never obtained in Eq.~(\ref{eq:heisprojeigenspace}) for any consecutive times $t_1 = q\tau$ and $t_2 = (q + 1)\tau$, we conclude that $M_\tau^k$ is a projector $\mathcal{H} \to V_\tau^1 \cap V_{2\tau}^1\cap \cdots \cap V_{k\tau}^1$, which in the limit of large $k$ converges to $\mathcal{H} \to \operatorname{span}\{\ket{\Lambda}\}$.

\begin{figure}
  \includegraphics[width=0.7\textwidth]{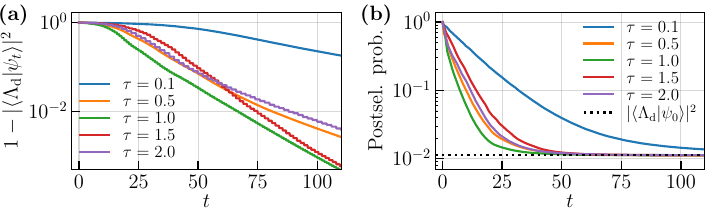}
  \caption{
    State preparation via stroboscopic $Z_1Z_{\dual 1}$ measurements separated by intervals $\tau$ (i.e., $k=\lfloor t/\tau\rfloor$ steps) on an $N=18$ dangler system.
    Here, $\ket{\psi_0} = \ket{00\ldots 0}$.
    The rate of exponential decrease of the preparation infidelity (a) and the corresponding rate of convergence of the overall postselection probability toward $|\braket{\Lambda_\text{d}}{\psi_0}|^2$ (b) depend on $\tau$: too frequent measurements impede information scrambling in the system due to quantum Zeno effect and result in slower convergence toward $\ket{\Lambda_\text{d}}$, whereas infrequent measurements are inefficient due to allowing the system to remain in an ``idle'' state in which the measurement-induced information has spread away from sites $1$ and $\dual 1$.
    Note that after a long enough series of successful measurements the infidelity of the state with respect to $\ket{\Lambda_\text{d}}$ becomes exponentially small in $k$ and unsuccessful measurements become extremely unlikely [i.e., the first factor in Eq.~(\ref{eq:postseldef}) becomes exponentially close to 1].
  }
  \label{fig:danglertau}
\end{figure}

This picture provides some insights into how state preparation dynamics depends on the interval between measurements.
The reason one cannot speed up state preparation via too frequent (small $\tau$) measurements is that subspaces $V_{q\tau}^1$ and $V_{(q+1)\tau}^1 = e^{iH\tau} V_{q\tau}^1$ almost perfectly overlap, which does not allow for significant leakage of any amplitude out of the time-dependent $\lambda=1$ sector; the system becomes semi-confined in its original subspace projected onto $V_\tau^1$ in a way reminiscent of the quantum Zeno effect (see Fig.~\ref{fig:danglertau}).

On the other hand, in the case of infrequent measurements (large generic $\tau$), $V_{q\tau}^1 \cap V_{(q+1)\tau}^1$ is, effectively, an overlap between two random subspaces in the full Hilbert space.
The average rate of amplitude leakage per measurement as $\tau$ gets large is expected to saturate at some purely geometric quantity, so infrequent measurements simply result in a poor ``duty cycle.''

Above, we reasoned in the operator picture without regard to states.
We now present a complementary point of view thinking more in terms of states.
Consider subspace orthogonal to $\ket{\Lambda}$, which we will denote $W$.
Both $P_{1\dual 1}^+$ and $H$ act within $W$, hence the same is true for $M_\tau$, and all the action below will be understood within $W$.
We want to see if generically $|\mu| < 1$ for all eigenvalues $\mu$ of $M_\tau$ in $W$.
Note that the operator $M_\tau$ is non-Hermitian, but we can still find its (in general complex) eigenvalues and bring it to the Jordan normal form using a similarity transformation.
Hence, if all $|\mu| < 1$, with finite Jordan blocks (true for finite-dimensional Hilbert spaces), then large powers of this operator decay to zero, i.e., $\lim_{k \to \infty} M_\tau^k W = 0$.
Note that there is no contradiction between $\| M_\tau \| = 1$ (which follows from earlier arguments) and eigenvalues of $M_\tau$ satisfying $|\mu| < 1$: $\| M_\tau \|$ is related to eigenvalues of $(M_\tau^\dagger M_\tau)^{1/2}$, which are not simply related to eigenvalues of $M_\tau$ for non-Hermitian $M_\tau$.
On the other hand, those earlier arguments do not impose conditions on $\|M_\tau^k\|$ in $W$, and the Jordan normal form arguments show that it will decay to zero for large $k$ if all $|\mu| < 1$.

Suppose $\mu$ is an eigenvalue of $M_\tau$ in $W$ and $\ket{\Phi} \in W$ is the corresponding eigenstate, $M_\tau \ket{\Phi} = \mu \ket{\Phi}$ (i.e., $\ket{\Phi}$ is a right eigenvector of $M_\tau$).
Since $\| M_\tau \ket{\Phi} \| \leq \|\ket{\Phi}\|$, we must have $|\mu| \leq 1$.
Suppose we have $|\mu| = 1$.  
Let us denote $\ket{\Theta} = e^{-iH\tau} \ket{\Phi}$.
We have $\| \ket{\Theta} \| = \| \ket{\Phi} \|$ and $P_{1\dual 1}^+ \ket{\Theta} = \mu \ket{\Phi}$, and since we have assumed $|\mu| = 1$ and $P_{1\dual 1}^+$ is a projector, we must have $P_{1\dual 1}^+ \ket{\Theta} = \ket{\Theta}$.
The eigenvector condition $M_\tau\ket{\Phi} = \mu \ket{\Phi}$ then implies $\ket{\Theta} = \mu\ket{\Phi}$, i.e., $e^{-iH\tau} \ket{\Phi} = \mu \ket{\Phi}$, which in turn means that $P_{1\dual 1}^+ \ket{\Phi} = \ket{\Phi}$.  
Thus, $\ket{\Phi}$ is a simultaneous eigenstate of $P_{1\dual 1}^+$ (with eigenvalue $+1$), and of $e^{-iH\tau}$.
Hence, if we can argue that $P_{1\dual 1}^+$ and $e^{-iH\tau}$ do not have simultaneous eigenstates in $W$, we obtain a contradiction, implying that $|\mu|$ cannot be equal to $1$ and hence must be smaller than $1$.

By the assumed uniqueness of $\ket{\Lambda}$, we know that $P_{1\dual 1}^+$ and $H$ cannot have simultaneous eigenstates in $W$.
This does not yet mean the same for $P_{1\dual 1}^+$ and $e^{-iH\tau}$, since we may have a situation where an eigenstate of $P_{1\dual 1}^+$ is, e.g., a superposition of two eigenstates of $H$ with distinct eigenvalues $\epsilon$ and $\epsilon'$ that happen to produce $e^{-i\epsilon\tau} = e^{-i\epsilon'\tau}$, and then this superposition is also an eigenstate of $e^{-iH\tau}$.
This situation clearly requires fine-tuning of $\tau$, as well as more special properties of $H$ like the above superposition of two eigenstates of $H$ being an eigenstate of $P_{1\dual 1}^+$.
Intuitively, even properties like the latter are unlikely for a generic chaotic Hamiltonian $H$; however, we do not need to assume this to see that for most values of $\tau$ we will have $e^{-i\epsilon\tau} \neq e^{-i\epsilon'\tau}$ for distinct eigenvalues $\epsilon \neq \epsilon'$, and the above essentially covers any adversarial situation leading to there being simultaneous eigenstates of $P_{1\dual 1}^+$ and $e^{-iH\tau}$ in $W$.
A more formal argument is as follows.
Suppose $\{\epsilon_n\}$ are distinct eigenvalues of $H$ with degeneracies $\{d_n\}$ and the corresponding eigenspaces $\{W_n\}$, $W = \oplus_n W_n$.
As long as we are dealing with finite sets (true for finite systems), clearly, for most choices of $\tau$ the corresponding phase values $\{e^{-i\tau\epsilon_n}\}$ will be distinct (possible $\tau$ violating this being measure zero).
These are then distinct eigenvalues of $e^{-iH\tau}$, and the corresponding eigenspaces are also uniquely fixed to be $\{W_n\}$.
Hence any eigenstate of $e^{-iH\tau}$ must be within one of the $W_n$'s and hence must also be an eigenstate of $H$.
Hence, any common eigenstate of $P_{1\dual 1}^+$ and $e^{-iH\tau}$ in $W$ is also an eigenstate of $H$, which is not allowed.

Note that in these more formal arguments we require $\tau$ to be away from some special fine-tuned values that depend on the system size, and we also considered the limit $k \to \infty$ while keeping the dimension of the Hilbert space fixed, which however becomes exponentially large with the system size.
While we cannot prove it, more suggestive earlier arguments and chaoticity of $H$ likely make the situation better producing reasonable convergence of $M_\tau^k W$ to zero for any generic $\tau$ and that does not require $k$ exponentially large in system size.
In the end, our numerical studies of the state preparation in the main text is the strong and most practical evidence for this.

\section{Measurement of the out-of-time-order correlation (OTOC) functions using \texorpdfstring{$\ket*{\widetilde\Lambda}_\mathcal{F}$}{Lambda-type} states}
\label{supp:otoc}
In this section we start by justifying the claim made in the main text that for short enough time $t < t_B$ (where $t_B$ is the ``butterfly'' time to be defined more precisely later) the two-point correlators $\ev{Z_iZ_{\dual i}} = \bra{\psi(t)} Z_iZ_{\dual i} \ket{\psi(t)}$ measured as a function of time $t$ in the quench experiment with $\ket{\psi(t=0)} = Z_L \ket{\Lambda}$ in Fig.~\suppref{fig:quench} are equivalent to the four-point out-of-time-order correlation (OTOC) function \suppcite{swingle2016measuring}
\begin{equation}
  F(t) = \ev{W_t^\dagger V^\dagger W_t V},
\end{equation}
where $W_t = Z_L(t)$ and $V = Z_i$, and the average is taken with respect to the infinite-temperature Gibbs ensemble of an OBC chain of size $L$.
We will then also show that the same protocol, when applied to identical decoupled OBC chains, allows measuring exact OTOCs for any time $t$. 
We will conclude the section with a short discussion on the more general applicability of our protocol to systems with different geometric configurations.

Denoting the unitary evolution by operator $U_t = e^{-iH_{\text{d}}t}$, the argument goes as follows:
\begin{subequations}
  \begin{align}
    \ev{Z_iZ_{\dual i}} &= \Tr\left(Z_iZ_{\dual i}U_tZ_L\ketbra{\Lambda_\text{d}}{\Lambda_\text{d}}Z_LU_t^\dagger\right) \\
    \label{eq:lambdainvu}
                        &=\Tr\left(Z_iZ_{\dual i}U_tZ_LU_t^\dagger\ketbra{\Lambda_\text{d}}{\Lambda_\text{d}}U_tZ_LU_t^\dagger\right) \\
                        &=\Tr\left(Z_iZ_{\dual i}Z_L(-t)\ketbra{\Lambda_\text{d}}{\Lambda_\text{d}}Z_L(-t)\right) \\
                        &=\Tr\left(Z_L(-t)Z_iZ_{\dual i}Z_L(-t)\ketbra{\Lambda_\text{d}}{\Lambda_\text{d}}\right) \\
    \label{eq:lambdainvzz}
                        &=\Tr\left(Z_L(-t)Z_iZ_{\dual i}Z_L(-t)Z_iZ_{\dual i}\ketbra{\Lambda_\text{d}}{\Lambda_\text{d}}\right) \\
    \label{eq:zibarcommute}
                        &\stackrel{t < t_B}{\simeq} \Tr\left(Z_L(-t)Z_iZ_L(-t)Z_i\ketbra{\Lambda_\text{d}}{\Lambda_\text{d}}\right) \\
                        &= \Tr_\mathcal{A} \Tr_\dualmathcal{A}\left(Z_L(-t)Z_iZ_L(-t)Z_i\ketbra{\Lambda_\text{d}}{\Lambda_\text{d}}\right) \\
                        &= \Tr_\mathcal{A} \sum_{f\in \mathcal{F}_L^{\text{OBC}}}\left(\mathbf{1}_\mathcal{A}\otimes \bra{f}_{\dual A}
                          \left[Z_L(-t)Z_iZ_L(-t)Z_i\ketbra{\Lambda_\text{d}}{\Lambda_\text{d}}\right]\mathbf{1}_\mathcal{A}\otimes \ket{f}_\dualmathcal{A}\right) \\
    \label{eq:fbarcommute}
                        &\stackrel{t < t_B}{\simeq} \Tr_\mathcal{A} \left(Z_L(-t)Z_iZ_L(-t)Z_i\sum_{f\in \mathcal{F}_L^{\text{OBC}}}\left[\mathbf{1}_\mathcal{A}\otimes \bra{f}_{\dual A}
                          \ketbra{\Lambda_\text{d}}{\Lambda_\text{d}}\mathbf{1}_\mathcal{A}\otimes\ket{f}_\dualmathcal{A}\right]\right) \\
                        &= \Tr_\mathcal{A} \left(Z_L(-t)Z_iZ_L(-t)Z_i\; \frac{1}{\left|\mathcal{F}_L^{\text{OBC}}\right|} \sum_{f\in \mathcal{F}_L^{\text{OBC}}}\ketbra{f}{f}\right) \\
                        &=\ev{Z_L(-t)Z_iZ_L(-t)Z_i}_{\{\ket{f}: f\in \mathcal{F}_L^{\text{OBC}}\}}.
  \end{align}
\end{subequations}
In lines labeled as Eqs.~(\ref{eq:lambdainvu}) and (\ref{eq:lambdainvzz}) we used the invariance of the density matrix $\ketbra{\Lambda_\text{d}}{\Lambda_\text{d}}$ under multiplication by $U_t$, $U_t^\dagger$  and $Z_iZ_{\dual i}$ from either left or right (since $\ket{\Lambda_\text{d}}$ is the eigenstate of all these operators with eigenvalue $+1$).
In lines labeled as Eqs.~(\ref{eq:zibarcommute}) and (\ref{eq:fbarcommute}) we assumed that the ``butterfly'' cone of the operator $Z_L(t)$ has not yet reached subsystem $\dualmathcal{A}$; thus the ``butterfly'' time $t_B$ approximately corresponds to the time when $\ev{Z_1Z_{\dual 1}}$ correlator starts decaying [see Fig.~\suppref{fig:quench}].
We also wrote schematically $\Tr$ as $\Tr_{\mathcal{A}} \Tr_\dualmathcal{A}$, where only $\Tr_\dualmathcal{A}$ is a true partial trace in the sense that it produces a RDM, for clarity; the outer $\Tr_\mathcal{A}$ is a full trace of the RDM defined on subsystem $\mathcal{A}$ that produces a scalar.
Finally, $\mathcal{F}_L^{\text{OBC}}$ is the set of Rydberg-blockaded bitstrings on the OBC chain of length $L$, and the last equation is infinite-temperature average over the corresponding ensemble.

\begin{figure}
  \subfloat{\label{fig:dhinfidelity}}
  \subfloat{\label{fig:dhpostsel}}
  \subfloat{\label{fig:dhquench}}
  \subfloat{\label{fig:otoccomp}}\par\nointerlineskip
  \includegraphics[width=\textwidth]{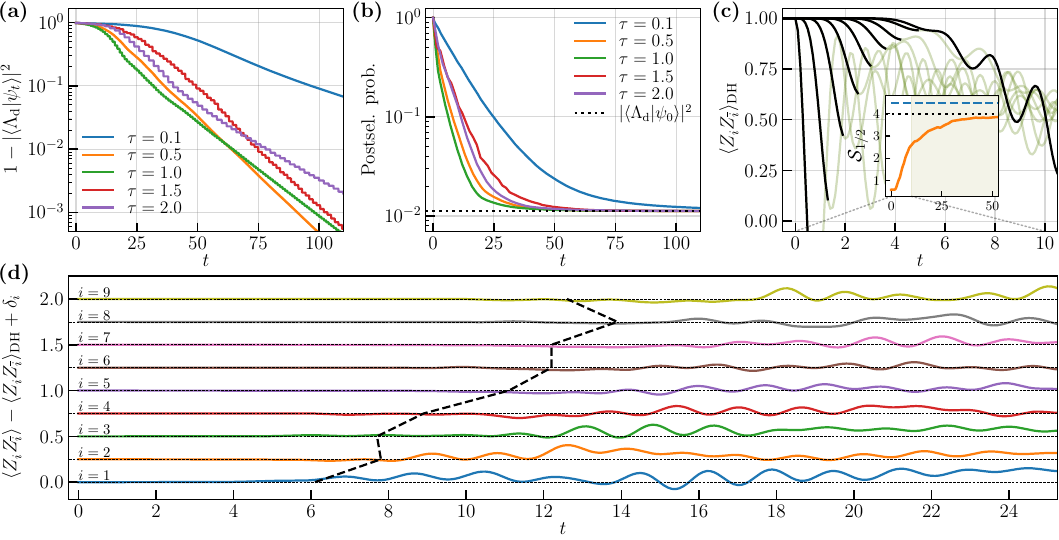}
  \caption{
    Numerical study of an $N=2L=18$ decoupled OBC chains system.
    (a),(b) State preparation via $Z_1Z_{\dual 1}$ measurements in a setting identical to that described in Fig.~\ref{fig:danglertau}.
    Note that the overall state preparation dynamics here is very similar to the example given in Fig.~\ref{fig:danglertau}; in fact, the rate of infidelity decay (a) is slightly faster in the decoupled system than in the dangler.
    (c) Quench experiment identical to that given in Fig.~\suppref{fig:quench} of the main text.
    The observables, as was anticipated in the discussion regarding the ``butterfly'' time $t_B$, are hardly distinguishable from those in the dangler system.
    Inset: no change in the half-system entanglement entropy of the standard bipartition (dashed line) occurs since the two subsystems are non-interacting under $H_\text{d}$; however, the entropy corresponding to the entanglement-minimizing bipartition of $\ket{\Lambda_\text{d}}$ (solid line) approaches the Page value (dotted line) as before.
    (d) Comparison between the approximate --- $\ev{Z_iZ_{\dual i}}$ --- and exact ---- $\ev{Z_iZ_{\dual i}}_\text{DH}$ --- OTOC proxies with shifts $\delta_i = (i-1)/4$ added for clarity.
    The dotted line approximately marks off the range of validity of the ``butterfly'' time approximation for each $i$ based on some heuristic detection of the onset of ``significant'' deviation between the two expectation values (the specifics of the heuristic are unimportant for this discussion).}
  \label{fig:dhprepotoc}
\end{figure}

The approximate $ZZ$ OTOC discussed above can be made exact.
By our construction in the main text, Fig.~\suppref{fig:dangler}, state $\ket{\Lambda_\text{d}}$ is also an eigenstate of the decoupled Hamiltonian $\widetilde H(G_\mathcal{A}\cup G_\dualmathcal{A}) = \widetilde H(G_\mathcal{A}) + \widetilde H(G_\dualmathcal{A})$, where $G_\mathcal{A}$ and $G_\dualmathcal{A}$ are isomorphic OBC chain graphs with $L$ vertices.
Per Theorem~\ref{thrm:discobc}, the same state preparation protocol as that discussed in the main text in context of the dangler system can be applied to a system consisting of two decoupled chains --- i.e., a system evolving under the Hamiltonian $\widetilde H(G_\mathcal{A}\cup G_\dualmathcal{A})$ [see Figs. \ref{fig:dhinfidelity} and \ref{fig:dhpostsel}].
Alternatively, if the experiment has the capability to decouple the subsystems $\mathcal{A}$ and $\dualmathcal{A}$ (e.g., by moving the two subsystems apart via optical tweezers) it may be possible switch from $H_{\text{d}}$ to the decoupled Hamiltonian $\widetilde H(G_\mathcal{A}) + \widetilde H(G_\dualmathcal{A})$ once the system is prepared in the eigenstate $\ket{\Lambda_\text{d}}$.
In either case, the same quench protocol as discussed earlier will allow to measure the exact ZZ OTOC on the subsystem $\mathcal{A}$ at any time $t$.

Since $[\widetilde H(G_\mathcal{A}), \widetilde H(G_\dualmathcal{A})] = 0$, the unitary evolution operator decouples into commuting terms: $U_t = U_\mathcal{A}U_\dualmathcal{A}$, where each term has support only on its corresponding subsystem and the time $t$ is kept implicit.
Then, following similar steps as before and simplifying, we have
\begin{equation}
  \begin{aligned}
  \ev{Z_iZ_{\dual i}} &= \Tr\left(Z_iZ_{\dual i}\cancel{U_\dualmathcal{A}}U_\mathcal{A}Z_LU_\mathcal{A}^\dagger\cancel{U_\dualmathcal{A}^\dagger}\ketbra{\Lambda_\text{d}}{\Lambda_\text{d}}\cancel{U_\dualmathcal{A}}U_\mathcal{A}Z_LU_\mathcal{A}^\dagger \cancel{U_\dualmathcal{A}^\dagger} \right) \\
                      &= \Tr\left(Z_iZ_{\dual i}Z_L(-t)_\mathcal{A}\ketbra{\Lambda_\text{d}}{\Lambda_\text{d}}Z_L(-t)_\mathcal{A} \right) \\
                      &= \Tr\left(Z_i\cancel{Z_{\dual i}}Z_L(-t)_\mathcal{A}Z_i\cancel{Z_{\dual i}}\ketbra{\Lambda_\text{d}}{\Lambda_\text{d}}Z_L(-t)_\mathcal{A} \right) \\
                      &= \Tr\left(Z_L(-t)_\mathcal{A}Z_iZ_L(-t)_\mathcal{A}Z_i\ketbra{\Lambda_\text{d}}{\Lambda_\text{d}} \right) \\
                      &=\ev{Z_L(-t)_\mathcal{A}Z_iZ_L(-t)_\mathcal{A}Z_i}_{\{\ket{f}: f\in \mathcal{F}_L^{\text{OBC}}\}},
  \end{aligned}
\end{equation}
where $Z_L(-t)_\mathcal{A}$ is operator $Z_L$ evolved under $\widetilde H(G_\mathcal{A})$ in the Heisenberg picture.
Since $\widetilde H(G_\mathcal{A})$ doesn't have support on the subsystem $\dualmathcal{A}$, $Z_L(-t)_\mathcal{A}$ never spreads outside of subsystem $\mathcal{A}$ (meaning it commutes with any operator with support on subsystem $\dualmathcal{A}$ only), and no approximations related to $t_B$ are necessary.
We show the exact OTOC measurements in Fig.~\ref{fig:dhquench} and compare them with the corresponding approximate measurement using the dangler Hamiltonian in Fig.~\ref{fig:otoccomp}.
Note that while approximation in Eq.~(\ref{eq:zibarcommute}) is valid for times up to $(L+\dual{i})/v_B$ where $v_B$ is the ``butterfly'' velocity, a more conservative estimate is assumed in Eq.~(\ref{eq:fbarcommute}); that being $t_B \sim L/v_B$ --- which appears to be around $t_B \sim 5$ for the cases shown in Figs.~\suppref{fig:quench} and \ref{fig:dhquench} --- since beyond this time $Z_L(-t)$ will be affected by the interaction term between the two half-systems of the dangler. Interestingly, this latter effect appears relatively smaller than expected, and for spins farther from the link connecting the two subsystems (i.e., from spins $1$ and $\dual 1$) the agreement between exact and approximate OTOCs in Fig.~\ref{fig:otoccomp} holds for significantly longer than $t_B$.

As was mentioned in the main text, there is strong numerical evidence that the dimension of the simultaneous eigenspace of $\widetilde H_\text{PXP}(G_\mathcal{C})$, where $G_\mathcal{C}$ is an arbitrary graph, and some $Z_iZ_{\dual i}$ on $\mathcal{R}$ is typically exactly equal to the number of distinct $\ket*{\widetilde\Lambda}_\mathcal{F}$ eigenstates whose pairing pattern contains pair $(i, \dual i)$.
This means that the state preparation protocol discussed in the main text can be applied to an arbitrary system given that it hosts a $\ket*{\widetilde\Lambda}_\mathcal{F}$ eigenstate with some known pairing pattern (see also Sec.~\ref{supp:mtau}).
Although our analysis of the approximate and exact OTOCs was done in the context of the dangler and disconnected OBC chains, no specific assumptions related to these systems had been made, so the same analysis applies to generic systems that host $\ket*{\widetilde\Lambda}_\mathcal{F}$ eigenstates.
Therefore, having prepared such a system in the desired $\ket*{\widetilde\Lambda}_\mathcal{F}$ state, one can execute the approximate or exact OTOC measurement protocol with essentially no difference from the two cases we addressed.
For example, one could study operator spreading on a two-dimensional OBC lattice.

\section{Effect of perturbations on the state preparation and OTOC protocols}
\label{supp:stability}
\begin{figure}
  \includegraphics[width=\textwidth]{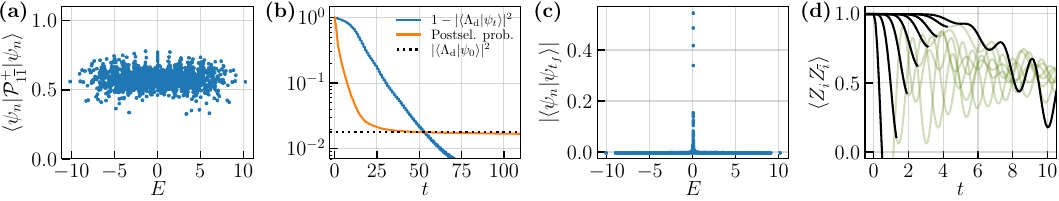}
  \caption{Perturbed dynamics with $\sigma = 0.01$ on a dangler system of size $N = 2L = 16$.
    (a) Squared projections of eigenstates onto the $+1$ eigenspace of $Z_1Z_{\dual 1}$. Note that the perturbation destroys the degenerate nullspace of $\widetilde H_\text{PXP}(G_\mathcal{C})$ as well as $\ket{\Lambda_\text{d}}$ as an eigenstate; there is no longer any simultaneous eigenstate of $\widetilde H_\text{PXP}(G_\mathcal{C})$ and $Z_1Z_{\dual 1}$.
    (b) State preparation protocol with the interval between measurements $\tau = 1.0$. Since the system has no simultaneous eigenstates of the full Hamiltonian and $Z_1Z_{\dual 1}$, the postselection probability doesn't saturate like in the clean case but instead undergoes a very slow decay following its initial fast drop toward $|\braket{\Lambda_\text{d}}{\psi_0}|^2$ slimilar to that in the clean case.
    The infidelity of the postselected state with respect to $\ket{\Lambda_\text{d}}$ decays similarly to the clean case, but is expected to eventually saturate at some finite value.
    (c) Overlap of the state prepared via postselection on the $+1$ outcomes of $Z_1Z_{\dual 1}$ with the eigenstates of the full Hamiltonian.
    The protocol drives the system into a superposition of many eigenstates with energies close to zero.
    (d) Quench under the perturbed Hamiltonian from the initial state $Z_L\ket{\psi_{t_f}}$, where $\ket{\psi_{t_f}}$ is the actual state prepared by the postselection protocol in (b).
    The observables are essentially indistinguishable from the clean case.
  }
  \label{fig:stab01}
\end{figure}

\begin{figure}
  \includegraphics[width=\textwidth]{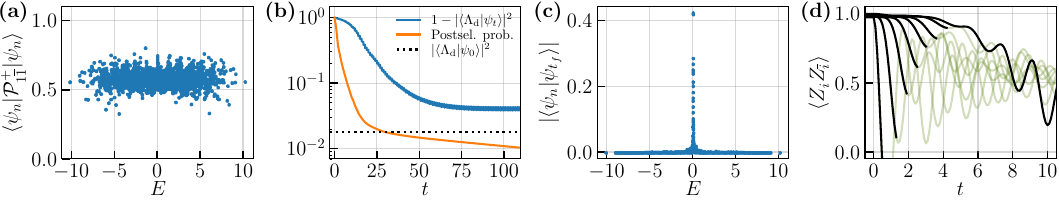}
  \caption{
    Perturbed dynamics with $\sigma = 0.02$.
    (a)-(d) Same as in Fig.~\ref{fig:stab01}.
    Note that the decay of the postselection probability becomes more noticeable here than in the $\sigma = 0.01$ case [Fig.~\ref{fig:stab01}].
    The infidelity of the postselected state with respect to $\ket{\Lambda_\text{d}}$ does not decay indefinitely like in the clean case, but saturates at a small finite value. This indicates a very large overlap of the postselected state with $\ket{\Lambda_\text{d}}$.
  }
  \label{fig:stab02}
\end{figure}

\begin{figure}
  \includegraphics[width=\textwidth]{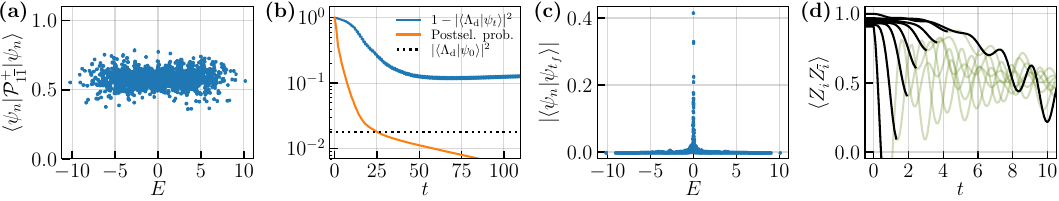}
  \caption{Perturbed dynamics with $\sigma = 0.03$.
    (a)-(d) Same as in Figs.~\ref{fig:stab01} and \ref{fig:stab02}.
    Here the decay of the postselection probability is even steeper than in the previous two examples. 
    Note, however, that the rate of decay of the postselection probability will depend on specific noise realization; i.e., it varies significantly depending on the random choices of the coefficients in Eq.~(\ref{eq:pert}).
    This means that in an experiment there will be implicit postselection of ``better'' noise realizations, which also come with lower state infidelities. 
    Here the effects of the perturbation on the OTOCs are already noticeable; however, the signal is still very close to that in the nearly unperturbed $\sigma = 0.01$ case.
  }
  \label{fig:stab03}
\end{figure}

Here we briefly investigate how the performance of the state preparation and OTOC protocols discussed in the main text and in Sec.~\ref{supp:otoc} of this Supplemental Material degrades when a perturbation is added to the Hamiltonian.
The intent of this section is to show that the dynamics involving $\ket*{\widetilde\Lambda}_\mathcal{F}$ states has a degree of robustness to small perturbations, even if such perturbations prevent $\ket*{\widetilde\Lambda}$ from being an exact eigenstate of the full Hamiltonian.
In this demonstration, we will perturb the dangler Hamiltonian $\widetilde H_\text{PXP}(G_\mathcal{C})$ with
\begin{equation}
  \label{eq:pert}
  \delta H = \sum_{i \in V} \delta_i Z_i,
\end{equation}
where $V$ is the set of vertices of $G_\mathcal{C}$, and the Zeeman fields $\delta_i$ are chosen randomly from a normal distribution with mean $\mu$ and variance $\sigma^2$, $\delta_i \sim \mathcal{N}(\mu,\,\sigma^{2})$.
We find that the results are qualitatively similar for randomized and uniform ($\sigma = 0$) perturbations.
The performance of the protocols gradually degrades as $\mu$ or $\sigma$ get larger.
In particular, in the numerical simulations presented in Figs.~\ref{fig:stab01}--\ref{fig:stab03} we set $\mu = 0$ and only varied $\sigma$ with the goal of modeling some experimentally undesirable non-uniform zero-mean Zeeman field; in these simulations, for each value of $\sigma$, we used a single fixed random sample of the $\delta_i$ coefficients in Eq.~(\ref{eq:pert}).
(In a more careful analysis the averages would be calculated over many such samples, but our goal here is to demonstrate the qualitative behavior only.)

\begin{figure}
  \includegraphics[width=\textwidth]{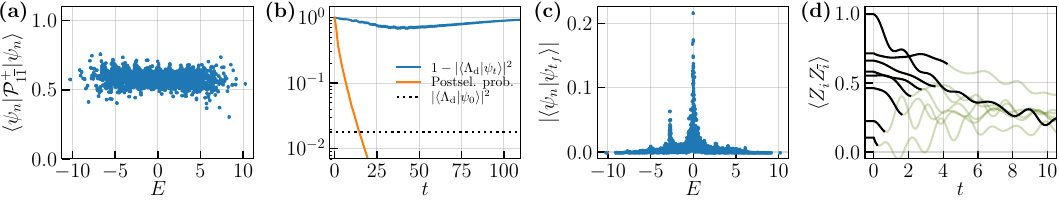}
    \caption{Perturbed dynamics with $\sigma = 0.1$.
    (a)-(d) Same as in Figs.~\ref{fig:stab01}--\ref{fig:stab03}.
    Strong perturbation breaks the state preparation protocol, but also drastically suppresses the postselection probability, which means situations like this are easily identifiable in experiments.
    As expected, the quench dynamics in (d) no longer closely resembles the clean case.
  }
  \label{fig:stab1}
\end{figure}

Even though eigenstate $\ket*{\Lambda_\text{d}}$ is destroyed by the perturbation, the simulations in Figs.~\ref{fig:stab01}--\ref{fig:stab03} clearly indicate that both the state preparation and quench dynamics retain most of the attributes found in the unperturbed case given the strength of the perturbation does not exceed some critical threshold [cf.~Fig.~\ref{fig:stab1}]. 
A more careful analysis of this threshold and its scaling with system size in a realistic experimental setting will be left to another study.

\section{State preparation protocol enhancements}
\label{supp:exp}
Here we briefly explore possible enhancements to the state preparation protocol discussed in the main text.
For concreteness, our examples will continue using the dangler geometry with $N=2L$ spins; however, everything will apply to other systems hosting $\ket*{\widetilde\Lambda}_\mathcal{F}$ eigenstates.

In general, the total number of measurements and time needed to prepare the system in the target state $\ket{\Lambda_\text{d}}$ with high fidelity can be reduced via a combination of two strategies: (1) increasing the overlap of the initial state with $\ket{\Lambda_\text{d}}$, and (2) performing measurements of more than a single pair of spins.

For an illustration of the first strategy, suppose it is possible to prepare Bell pairs
\begin{equation}
  \ket{\Phi^-} = \frac{1}{\sqrt{2}}\left(\ket{00} - \ket{11}\right).
\end{equation}
Then, assuming $L$ is even, consider state
\begin{equation}
\label{eq:phi0initial}
  \ket{\phi_0} = \bigotimes_{i=1}^{L/2}\left(\ket{\Phi^-}_{2i-1,\overline{2i-1}}\ket{00}_{2i,\overline{2i}}\right),
\end{equation}
 which can be prepared via two-qubit gates without violating the Rydberg blockade. Its overlap with $\ket{\Lambda_\text{d}}$ is
\begin{equation}
  \braket{\phi_0}{\Lambda_\text{d}} = \sqrt{\frac{2^{L/2}}{F_{L+2}}} = 2^{L/4} \braket{\psi_0}{\Lambda_\text{d}},
\end{equation}
where $\ket{\psi_0}=\ket{00\ldots 0}$ is the initial state used in the main text.
This is already a significant enhancement, strongly suppressing the exponential decay with $L$ [comparing $F_{L+2} = C \phi^L \approx C (1.618)^L$ vs $F_{L+2}/2^{L/2} = C (\phi/\sqrt{2})^L \approx C (1.144)^L$].
Note that in an actual experiment one may be able to produce initial states that have even greater overlaps with $\ket{\Lambda_\text{d}}$ by entangling all pairs of spins (as opposed to only half of them as in $\ket{\phi_0}$) and relying on the natural Rydberg blockade.

As for the second strategy, consider postselection on the $+1$ outcomes of a set of independent and simultaneous projective measurements $\mathcal{P}^{+}_{i\dual{i}} = \ketbra{00}_{i\dual i} + \ketbra{11}_{i\dual i}, i \in \Omega$, where $\Omega \subseteq \{1,\ldots,L\}$. This corresponds to the following generalized composite evolution and projective measurement operator $M'_\tau$:
\begin{equation}
  \label{eq:mtaumod}
  M'_\tau = \left(\prod_{i\in \Omega}\mathcal{P}_{i\dual{i}}\right)e^{-iH_\text{d}\tau}.
\end{equation}
Setting $\Omega = \{1\}$ reduces $M'_\tau$ to $M_\tau$ in the main text, whereas setting $\Omega$ to any other non-empty subset of $\{1,\ldots,L\}$ results in an extended and generally more efficient operator that is also somewhat less sensitive to the speed of information scrambling in the system (i.e., the optimal frequency of measurements typically increases if one adds additional spin pairs to the protocol).

\begin{figure}
  \includegraphics[width=0.667\textwidth]{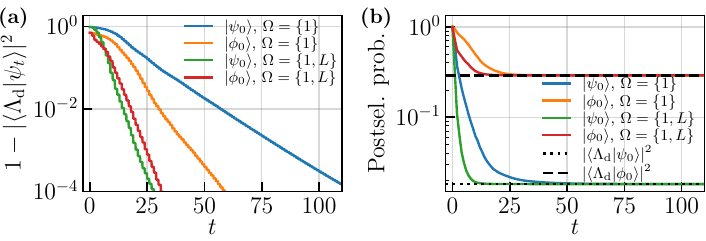}
  \caption{Dependence of the infidelity (a) and postselection probability (b) on the choice of the initial state and number of simultaneously measured spin pairs in an $N=2L=16$ dangler system with the interval between measurements $\tau = 1.0$.
  Here $\ket{\psi_0}=\ket{00\ldots 0}$, $\ket{\phi_0}$ is the state in Eq.~(\ref{eq:phi0initial}), and the composite evolution and projective measurement operator is given by Eq.~(\ref{eq:mtaumod}) with the sets $\Omega$ as specified in the legend.}
  \label{fig:prepenh}
\end{figure}

The two strategies discussed above are demonstrated in Fig.~\ref{fig:prepenh}. Note that while the second strategy reduces the number of measurements needed for preparing the target state with desired fidelity, it cannot increase the overall postselection probability, which depends solely on the overlap of the initial state with $\ket{\Lambda_\text{d}}$.